\documentstyle[psfig,12pt]{article}
\title{New CP-violating parameters in cascade decays}
\author{A.\ Amorim$^{a,b}$, M{\'a}rio G.\ Santos$^a$ and
Jo{\~a}o P.\ Silva$^{a,c}$\\
\\
\small $^a$ Centro de F{\'\i}sica Nuclear da Universidade de Lisboa\\
\small Avenida\ Professor\ Gama Pinto, 2,\\
\small 1699 Lisboa Codex, Portugal\\
\small \\
\small $^b$ Departamento de F{\'\i}sica\\
\small Faculdade de Ci{\^e}ncias da Universidade de Lisboa\\
\small Campo Grande, 1700 Lisboa, Portugal\\
\small \\
\small $^c$ Centro de F{\'\i}sica\\
\small Instituto Superior de Engenharia de Lisboa\\
\small 1900 Lisboa, Portugal}
\begin{document}
\maketitle
\begin{abstract}
We consider decay chains of the type
$P \rightarrow M + \cdots \rightarrow f + \cdots$,
where $M$ is a neutral meson that may mix with its antiparticle
$\overline{M}$,
before decaying into the final state $f$.
$P$ may be either a heavier neutral meson or a charged meson.
We perform a rephasing-invariant analysis of the quantities that show up
in such cascade decays.
If the decay $P \rightarrow M + \cdots$
(or the decay $P \rightarrow \overline{M} + \cdots$) is forbidden,
we find the usual $\lambda_f$ parameters describing the interference
between the mixing of a neutral meson system and the decay
{\em from that system}
into the final state $f$. 
However,
when both the $P \rightarrow M + \cdots$ and
$P \rightarrow \overline{M} + \cdots$ decays are allowed,
we find a new class of rephasing-invariant parameters,
$\xi_i$,
that measure the interference between the mixing of a neutral
meson system and the decay from the initial state
{\em into that system}.
We show that the quantities $\lambda_f$ and $\xi_i$ are necessary
and sufficient to describe all the interference effects present in the most
general cascade decay.
We discuss the various cascade decays in turn,
highlighting the special features of each one.
\end{abstract}


\section{Introduction}

The particle-antiparticle neutral meson systems have long been identified
as an ideal setting to search for CP violation (CPV).
This effect was discovered in 1964 in the neutral kaon system \cite{Chr64},
and it is believed that this will soon be complemented with
measurements of CPV in the decays of neutral $B$ mesons.

The CP violation found in the kaon system and the null results found
elsewhere can be accommodated in the Standard Model (SM),
through an irremovable complex phase in the Cabibbo-Kobayashi-Maskawa
quark mixing matrix \cite{CKM}.
The major goal of the upcoming $B$-factory experiments is to submit
this explanation to stringent tests \cite{Sanda}.

In 1989, Azimov \cite{Azimov} pointed out that additional tests could be
performed by looking for
$B_d \rightarrow J/\psi K \rightarrow J/\psi [f]_K$ `cascade decays',
involving the neutral $B_d$ and neutral kaon systems in succession.
This idea has been followed by Dass and Sarma \cite{Dass}.
Recently, 
cascade decays have received renewed attention.
Kayser and Stodolsky have developed a matrix method
to deal with cascade decays \cite{Kay96};
Kayser has shown that one may use
$B_d \rightarrow J/\psi K \rightarrow J/\psi [\pi l \nu]_K$
decays to get at $\cos 2 \beta$ \cite{Kay97};
and Azimov and Dunietz pointed out that one may use
$B_s \rightarrow J/\psi K \rightarrow J/\psi [f]_K$
decays to measure $\Delta m_{B_s}$,
even if $\Delta m_{B_s}$ turns out to be too large to be determined
experimentally from the direct decays of the $B_s$ mesons \cite{AziDun}.

In all these cases, an initial $B^0$ meson\footnote{We will
take $B^0$ to stand for both $B^0_d$ and $B^0_s$.
Moreover,
$K$ refers to a generic combination of $K^0$ and $\overline{K^0}$,
and similarly for $D$, $B_d$ and $B_s$.}
can only decay to one of the kaon's flavour eigenstates.
To leading order in the SM,
the decays $B^0_d \rightarrow \overline{K^0} + X$ and
$B^0_s \rightarrow K^0 + X$, and the respective CP conjugate decays
are forbidden.
In this article we will consider a more general situation,
in which a given meson can decay into both flavour eigenstates
of a lighter neutral meson system.
This situation is the one relevant for the decays
\begin{eqnarray}
\label{decays}
B^\pm \rightarrow D + X^\pm \rightarrow [f]_D + X^\pm,
& \hspace{5mm} &
B \rightarrow D + X \rightarrow [f]_D + X,
\nonumber\\
D^\pm \rightarrow K + X^\pm \rightarrow [f]_K + X^\pm,
& \hspace{5mm} &
D \rightarrow K + X \rightarrow [f]_K + X.
\end{eqnarray}
Decays of the first type have been discussed by Meca and Silva \cite{Silva}
in the context of uncovering new physics effects in the
$D^0 - \overline{D^0}$ system.
Here we show that all these decays depend on a new type of CP-violating
observables, over and above those required for the description of
direct decays of neutral mesons.

In section 2 we describe all the quantities involved in cascade decays
of the type $P \rightarrow M + X \rightarrow [f]_M + X$,
where $P$ and $M$ describe two neutral meson systems,
and $[f]_M$ describes a set of particles into which $M$ can decay,
and which have an invariant mass equal to that of the
$M^0 - \overline{M^0}$ system.
This analysis is performed by noting that all physical quantities must
remain invariant under a rephasing of the kets in the problem.
In section 3 we develop the conditions for CP invariance,
thus identifying all the CP-violating rephasing-invariant quantities
involved in cascade decays.
Naturally,
we obtain the usual CP-violating observables.
These describe CPV in the mixing of the neutral meson systems;
CPV in the decays;
and CPV in the interference between the mixing of the neutral meson system
and the decay {\em from that system} into the final state.
However,
when both the $P \rightarrow M + \cdots$ and
$P \rightarrow \overline{M} + \cdots$ decays are allowed,
we find a new class of rephasing invariant quantities that measure
the interference between the mixing of the intermediate neutral meson system
$M$,
and the decay from the initial state $P$ {\em into that system}.
These quantities are used in section 4 to develop the formulae describing
a generic cascade decay.
Section 5 contains a classification of cascade decays and an extended
discussion of the special features of each case.
We draw our conclusions in section 6.

\section{Rephasing-invariant quantities}

Let us consider a decay of the type
\begin{equation}
P \rightarrow M + X \rightarrow [f]_M + X,
\end{equation}
where both $P$ and $M$ belong to neutral meson systems.
In general, the amplitude for the overall decay chain involves the
amplitudes of the primary decays
\begin{eqnarray}
A_{P^0 \rightarrow M^0} \equiv
\langle M^0 X | T | P^0 \rangle\ ,
& \hspace{5mm} &
A_{\overline{P^0} \rightarrow M^0} \equiv
\langle M^0 X | T | \overline{P^0} \rangle\ ,
\nonumber\\
A_{P^0 \rightarrow \overline{M^0}} \equiv
\langle \overline{M^0} X | T | P^0 \rangle\ ,
& \hspace{5mm} &
A_{\overline{P^0} \rightarrow \overline{M^0}} \equiv
\langle \overline{M^0} X | T | \overline{P^0} \rangle\ ,
\end{eqnarray}
followed by the amplitudes of the secondary decay,
\begin{equation}
A_{M^0 \rightarrow f} \equiv
\langle f | T | M^0 \rangle\ ,
\hspace{5mm}
A_{\overline{M^0} \rightarrow f} \equiv
\langle f | T | \overline{M^0} \rangle\ .
\end{equation}
Since there is mixing in the neutral $P^0 - \overline P^0$ system,
the decay rate will also involve the parameter $q_P/p_P$ 
which arises in the transformation from the flavour eigenstates
into the mass eigenstates:
\begin{eqnarray}
| P_H \rangle &=& p_P | P^0 \rangle + q_P | \overline{P^0} \rangle,
\nonumber\\
| P_L \rangle &=& p_P | P^0 \rangle - q_P | \overline{P^0} \rangle.
\end{eqnarray}
Here, $P_H$ and $P_L$ refer to the heavy and light mass eigenstates,
respectively.
Similarly,
the $M^0 - \overline{M^0}$ mixing enters in the expressions through
the parameter $q_M/p_M$.
For completeness,
we describe in appendix~\ref{appendix:evolution} the time evolution
of a generic neutral meson system.

In quantum mechanics,
all kets may be rephased at will,\footnote{The rephasing-invariant
analysis of cascade decays presented in this section is inspired
in the analysis of direct decays performed in Ref.~\cite{Luis}.}
\begin{equation}
\label{eq:rephazing-kets}
\begin{array}{rclcrcl}
| P^0 \rangle & \rightarrow & e^{i\gamma_P} |P^0 \rangle\ ,
&\hspace{5mm}&
|\overline{P^0}\rangle & \rightarrow &
e^{i\overline{\gamma}_P} |\overline{P^0} \rangle\ ,
\\
| M^0 \rangle & \rightarrow & e^{i\gamma_M} | M^0 \rangle\ ,
&\hspace{5mm}&
| \overline{M^0} \rangle & \rightarrow &
e^{i\overline{\gamma}_M} | \overline{M^0} \rangle\ ,
\\
| f \rangle & \rightarrow & e^{i\gamma_f} | f \rangle\ ,
&\hspace{5mm}&
| \bar f \rangle & \rightarrow &
e^{i\overline{\gamma}_f} | \bar f \rangle\ .
\end{array}
\end{equation}
Under this rephasing,
the amplitudes and mixing parameters also change,
according to
\begin{equation}
\label{eq:rephazing-quantities}
\begin{array}{rclcrcl}
\frac{q_P}{p_P} & \rightarrow & 
e^{i(\gamma_P - \overline{\gamma}_P)} \frac{q_P}{p_P}\ , &
&
\frac{q_M}{p_M} & \rightarrow &
e^{i(\gamma_M - \overline{\gamma}_M)} \frac{q_M}{p_M}\ ,
\\*[2mm]
A_{P^0 \rightarrow M^0} & \rightarrow &
e^{i(\gamma_P-\gamma_M)} A_{P^0 \rightarrow M^0}\ , &
&
A_{\overline{P^0} \rightarrow M^0} & \rightarrow &
e^{i(\overline{\gamma}_P-\gamma_M)} A_{\overline{P^0} \rightarrow M^0}\ ,
\\*[2mm]
A_{P^0 \rightarrow \overline{M^0}} & \rightarrow &
e^{i(\gamma_P-\overline{\gamma}_M)} A_{P^0 \rightarrow \overline{M^0}}\ , &
&
A_{\overline{P^0} \rightarrow \overline{M^0}} & \rightarrow &
e^{i(\overline{\gamma}_P-\overline{\gamma}_M)}
A_{\overline{P^0} \rightarrow \overline{M^0}}\ ,
\\*[2mm]
A_{M^0 \rightarrow f} & \rightarrow &
e^{i(\gamma_M - \gamma_f)} A_{M^0 \rightarrow f}\ , &
&
A_{\overline{M^0} \rightarrow f} & \rightarrow &
e^{i(\overline{\gamma}_M - \gamma_f)} A_{\overline{M^0} \rightarrow f}\ ,
\\*[2mm]
A_{M^0 \rightarrow \bar f} & \rightarrow &
e^{i(\gamma_M - \overline{\gamma}_f)} A_{M^0 \rightarrow \bar f}\ , &
&
A_{\overline{M^0} \rightarrow \bar f} & \rightarrow &
e^{i(\overline{\gamma}_M - \overline{\gamma}_f)}
A_{\overline{M^0} \rightarrow \bar f}\ .
\end{array}
\end{equation}
Only those quantities which are invariant under this redefinition
may have physical meaning.
Clearly,
the magnitudes of all the quantities in Eq.~\ref{eq:rephazing-quantities}
satisfy this condition.

In addition,
we find some rephasing-invariant quantities which arise from
the clash between the phases in the mixing and the phases in the decay
amplitudes:
\begin{eqnarray}
\lambda_{P \rightarrow M^0} \equiv \frac{q_P}{p_P}
\frac{A_{\overline{P^0} \rightarrow M^0}}{A_{P^0 \rightarrow M^0}}\ ,
& \hspace{5mm} &
\lambda_{P \rightarrow \overline{M^0}} \equiv \frac{q_P}{p_P}
\frac{A_{\overline{P^0} \rightarrow \overline{M^0}}}{
A_{P^0 \rightarrow \overline{M^0}}}\ ,
\label{def:int-1}
\\ 
\lambda_{M \rightarrow f} \equiv \frac{q_M}{p_M}
\frac{A_{\overline{M^0} \rightarrow f}}{A_{M^0 \rightarrow f}}\ ,
& \hspace{5mm} &
\lambda_{M \rightarrow \bar f} \equiv \frac{q_M}{p_M}
\frac{A_{\overline{M^0} \rightarrow \bar f}}{A_{M^0 \rightarrow \bar f}}\ ,
\label{def:int-2}
\\
\xi_{P^0 \rightarrow M} \equiv
\frac{A_{P^0 \rightarrow \overline{M^0}}}{A_{P^0 \rightarrow M^0}}
\frac{p_M}{q_M}\ ,
& \hspace{5mm} &
\xi_{\overline{P^0} \rightarrow M} \equiv
\frac{A_{\overline{P^0} \rightarrow \overline{M^0}}}{
A_{\overline{P^0} \rightarrow M^0}}
\frac{p_M}{q_M}\ .
\label{def:int-3}
\end{eqnarray}
The quantities in Eq.~\ref{def:int-1} are well known.
They describe the interference between the mixing in the
$P^0 - \overline{P^0}$ system and the decay {\em from that system}
into the flavour specific meson $M^0$ or $\overline{M^0}$,
respectively.
Similarly,
the parameters in Eq.~\ref{def:int-2} describe the interference between
the mixing in the $M^0 - \overline{M^0}$ system and the decay
{\em from that system}
into the final state $f$ and $\bar f$, respectively.
On the contrary,
the parameters in Eq.~\ref{def:int-3} describe a new type of interference:
the interference between the mixing in the $M^0 - \overline{M^0}$ system
and the decay {\em into that system} (originating from $P^0$,
in the case of $\xi_{P^0 \rightarrow M}$,
or from $\overline{P^0}$,
in the case of $\xi_{\overline{P^0} \rightarrow M}$).

We notice that the parameters described thus far are not all independent.
In particular
\begin{equation}
\label{eq:relate}
\lambda_{P \rightarrow M^0}
\xi_{\overline{P^0} \rightarrow M}
=
\lambda_{P \rightarrow \overline{M^0}}
\xi_{P^0 \rightarrow M}.
\end{equation}
This means that,
of the six phases contained in Eqs.~\ref{def:int-1},
\ref{def:int-2},
and \ref{def:int-3},
only five are independent.
This is obvious from Eqs.~\ref{eq:rephazing-quantities}
which contain ten quantities and five rephasings.

\section{Conditions for CP-invariance}

In order to study CP violation we first consider the consequences of CP
invariance.
If CP is a good symmetry,
then there exist three phases $\alpha_P$, $\alpha_M$ and $\alpha_f$,
and two CP eigenvalues $\eta_P= \pm 1$ and $\eta_M = \pm 1$,
such that:
\begin{eqnarray}
{\cal CP} | P^0 \rangle & = & e^{i\alpha_P} | \overline{P^0} \rangle\ ,
\nonumber\\
{\cal CP} | M^0 \rangle & = & e^{i\alpha_M} | \overline{M^0} \rangle\ ,
\nonumber\\
{\cal CP} | f \rangle & = & e^{i\alpha_f} | \bar f \rangle\ ,
\label{eq:CP1}
\end{eqnarray}
and
\begin{equation}
\begin{array}{rclcrcl}
{\cal CP} | P_H \rangle &=& \eta_P | P_H \rangle\ , &
& 
{\cal CP} | P_L \rangle &=& - \eta_P | P_L \rangle\ ,
\\
{\cal CP} | M_H \rangle &=& \eta_M | M_H \rangle\ , &
&
{\cal CP} | M_L \rangle &=& - \eta_M | M_L \rangle\ .
\end{array}
\label{eq:CP2}
\end{equation}
Notice that, 
since we define the sign of $\Delta m$ to be positive,
it is the value of $\eta_P$ and of $\eta_M$ that must be taken from experiment.
For example, we know experimentally that,
if the small CP violation in the neutral kaon system were absent,
then the heavy kaon would coincide with the long-lived kaon,
which would be CP odd.
Therefore,
we must take $\eta_K = -1$ whenever we neglect CP violation in the
neutral kaon system.
For the other neutral meson systems,
these parameters must still be determined experimentally.
In the meantime,
one sometimes uses the SM predictions for these parameters.
For instance,
the SM prediction for the $B^0_s - \overline{B^0_s}$ system is that the
heavy state should be mostly CP odd \cite{Dun95}.
In this case,
neglecting the CP violation in the $B^0_s - \overline{B^0_s}$
system leads to $\eta_{B_s} = -1$.

Moreover,
the CP transformation of the composite intermediate state is
\begin{equation}
{\cal CP} | X M^0 \rangle  =  \eta_X e^{i\alpha_M} 
| \overline{X}\,  \overline{M^0} \rangle\ .
\label{eq:CP3}
\end{equation}
Here $\eta_X$ contains any relevant CP transformation of the state $X$,
as well as the parity corresponding to the relative angular momentum between
$X$ and $M$.
For example,
in the decays $B \rightarrow J/\psi K$,
$\eta_X = - 1$ because $J/\psi$ is CP even,
while the $J/\psi$ and the kaon have a relative $L=1$ orbital
angular momentum \cite{Thank you}.

As a result of Eqs.~\ref{eq:CP1}, \ref{eq:CP2} and \ref{eq:CP3},
the CP invariance conditions are
\begin{equation}
\frac{q_P}{p_P} = \eta_P e^{i \alpha_P}\ ,
\hspace{10mm}
\frac{q_M}{p_M} = \eta_M e^{i \alpha_M}\ ,
\end{equation}
and
\begin{equation}
\label{need this}
\begin{array}{rclcrcl}
A_{P^0 \rightarrow M^0} & = &
\eta_X
e^{i(\alpha_P - \alpha_M)} A_{\overline{P^0} \rightarrow \overline{M^0}}\ , &
&
A_{P^0 \rightarrow \overline{M^0}} & = &
\eta_X
e^{i(\alpha_P + \alpha_M)} A_{\overline{P^0} \rightarrow M^0}\ ,
\\*[2mm]
A_{M^0 \rightarrow f} & = &
e^{i(\alpha_M - \alpha_f)} A_{\overline{M^0} \rightarrow \bar f}\ , &
&
A_{M^0 \rightarrow \bar f} & = &
e^{i(\alpha_M + \alpha_f)} A_{\overline{M^0} \rightarrow f}\ .
\end{array}
\end{equation}
Therefore, if CP is conserved we have
\begin{equation}
\begin{array}{rclcrcl}
\left| \frac{q_P}{p_P} \right| & = & 1 &
&
\left| \frac{q_M}{p_M} \right| & = & 1
\\*[2mm]
\left| A_{P^0 \rightarrow M^0} \right| & = &
\left| A_{\overline{P^0} \rightarrow \overline{M^0}} \right| &
&
\left| A_{P^0 \rightarrow \overline{M^0}} \right| & = &
\left| A_{\overline{P^0} \rightarrow M^0} \right|
\\*[2mm]
\left| A_{M^0 \rightarrow f}  \right|& = &
\left| A_{\overline{M^0} \rightarrow \bar f} \right|&
&
\left| A_{M^0 \rightarrow \bar f} \right| & = &
\left| A_{\overline{M^0} \rightarrow f} \right|\ .
\end{array}
\end{equation}
Moreover, the parameters describing interference CP violation become
related by
\begin{eqnarray}
\lambda_{P \rightarrow M^0}\, \lambda_{P \rightarrow \overline{M^0}}
& = & 1
\nonumber\\
\lambda_{M \rightarrow f}\, \lambda_{M \rightarrow \bar f}
& = & 1
\nonumber\\
\xi_{P^0 \rightarrow M}\, \xi_{\overline{P^0} \rightarrow M}
& = & 1
\nonumber\\
\lambda_{P \rightarrow M^0}\, \xi_{\overline{P^0} \rightarrow M}
=
\lambda_{P \rightarrow \overline{M^0}}\, \xi_{P^0 \rightarrow M}
& = & \eta_X \eta_P \eta_M.
\end{eqnarray}
We have used Eq.~\ref{eq:relate} on the last line.
One may use the relations above to develop more complicated conditions for 
CP invariance, such as
\begin{equation}
A_{\overline{P^0} \rightarrow \overline{M^0}}
A_{\overline{M^0} \rightarrow f}
A^\ast_{\overline{P^0} \rightarrow M^0}
A^\ast_{M^0 \rightarrow f}
=
A_{P^0 \rightarrow M^0}
A_{M^0 \rightarrow \bar f}
A^\ast_{P^0 \rightarrow \overline{M^0}}
A^\ast_{\overline{M^0} \rightarrow \bar f}.
\end{equation}

A very important special case occurs when $f$ is a CP eigenstate.
Then $\eta_f \equiv e^{i \alpha_f} = \pm 1$,
and the conditions for CP invariance become
\begin{equation}
\label{CP inv with CP eigen}
\left| A_{M^0 \rightarrow f} \right| = 
\left| A_{\overline{M^0} \rightarrow f} \right|
\ \ \mbox{and }\ \ 
\lambda_{M \rightarrow f} = \eta_M \eta_f.
\end{equation}

We have found the usual types of CP violation:
$|q/p| - 1$ probes CP violation in the mixing of neutral mesons;
$|A_{i \rightarrow f}| - |A_{\bar i \rightarrow \bar f}|$ probes
CP violation in the decays;
while $\arg \lambda_{i \rightarrow f} + \arg \lambda_{i \rightarrow \bar f}$
measures CP violation arising in the interference between the
mixing of the initial neutral meson system,
$i$,
and the decay {\em from that system} into the final states $f$ and $\bar f$.
If $f$ is a CP eigenstate,
then the CP-violating observable
$\arg \lambda_{i \rightarrow f} + \arg \lambda_{i \rightarrow \bar f}$
is related to $\mbox{Im} \lambda_{i \rightarrow f}$.
In addition,
we have discovered a new type of CP-violating observables given by
$\arg \xi_{P^0 \rightarrow M} + \arg \xi_{\overline{P^0} \rightarrow M}$.
These observables measure the CP violation which arises from the interference
between the mixing in the $M^0 - \overline{M^0}$ system and the
decay from the initial state ($P^0$ and $\overline{P^0}$)
{\em into that system}.

\section{The decay rate for a generic cascade decay}

\subsection{Decay amplitudes}

Let us consider a neutral meson $P^0$ that decays into an intermediate
state containing a combination of $M^0$ and $\overline{M^0}$ at time
$t_P$, measured in the rest-frame of $P^0$.
The resulting combination of $M^0$ and $\overline{M^0}$ will evolve in time
and decay into the final state $f$ at time $t_M$,
measured in its rest-frame.
The decay amplitude for this process is given by
\begin{eqnarray}
A \left[ P^0 \stackrel{t_P}{\rightarrow} M 
\stackrel{t_M}{\rightarrow} f \right] 
& = &
g_+^P(t_P) g_+^M(t_M)
\left[ 
A_{P^0 \rightarrow M^0} A_{M^0 \rightarrow f} + 
A_{P^0 \rightarrow \overline{M^0}} A_{\overline{M^0} \rightarrow f} 
\right]
\nonumber\\
& + &
g_+^P(t_P) g_-^M(t_M)
\left[
\frac{q_M}{p_M} A_{P^0 \rightarrow M^0} A_{\overline{M^0} \rightarrow f} +
\frac{p_M}{q_M} A_{P^0 \rightarrow \overline{M^0}} A_{M^0 \rightarrow f}
\right]
\nonumber\\
& + &
g_-^P(t_P) g_-^M(t_M) \frac{q_P}{p_P}
\left[
\frac{p_M}{q_M} 
A_{\overline{P^0} \rightarrow \overline{M^0}} A_{M^0 \rightarrow f} +
\frac{q_M}{p_M}
A_{\overline{P^0} \rightarrow M^0} A_{\overline{M^0} \rightarrow f}
\right]
\nonumber\\
& + &
g_-^P(t_P) g_+^M(t_M) \frac{q_P}{p_P}
\left[ 
A_{\overline{P^0} \rightarrow \overline{M^0}} A_{\overline{M^0} \rightarrow f} 
+
A_{\overline{P^0} \rightarrow M^0} A_{M^0 \rightarrow f}
\right].
\end{eqnarray}
This result is easiest to derive with the help of the evolution
diagram presented in Fig.~1.
\begin{figure}
\centerline{\psfig{figure=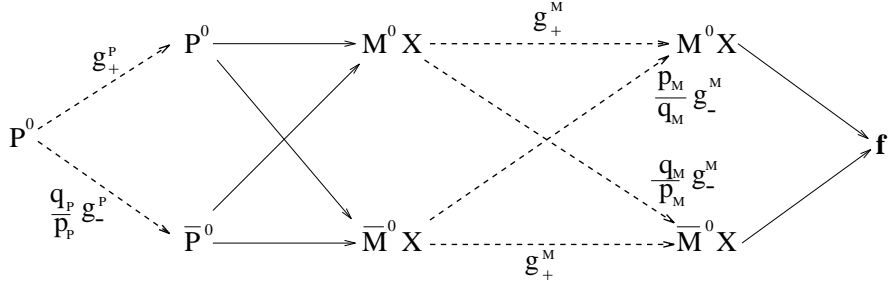,height=1.5in}}
\caption{The most general $P \rightarrow M X \rightarrow [f]_M X$
cascade decay.
\label{fig:1}}
\end{figure}
Henceforth we shall not show the explicit time dependence of the
$g$-functions.

The decay amplitude may also be written as
\begin{eqnarray}
\label{amplitude:P}
A \left[ P^0 \stackrel{t_P}{\rightarrow} M 
\stackrel{t_M}{\rightarrow} f \right]
&=&
\left( 
A_{P^0 \rightarrow M^0} A_{M^0 \rightarrow f} + 
A_{P^0 \rightarrow \overline{M^0}} A_{\overline{M^0} \rightarrow f}
\right) \times
\nonumber\\
& &
\hspace{2mm}
\left[
g_+^P g_+^M +
\chi_1\,  g_+^P g_-^M +
\chi_2\, g_-^P g_-^M +
\chi_3 g_-^P g_+^M
\right]\ ,
\end{eqnarray}
where we have used
\begin{equation}
\begin{array}{rclcl}
\chi_1 & \equiv &
\frac{A_{P^0 \rightarrow M^0} \frac{q_M}{p_M} 
A_{\overline{M^0} \rightarrow f} + 
A_{P^0 \rightarrow \overline{M^0}} \frac{p_M}{q_M}
A_{M^0 \rightarrow f}}{
A_{P^0 \rightarrow M^0} A_{M^0 \rightarrow f} + 
A_{P^0 \rightarrow \overline{M^0}} A_{\overline{M^0} \rightarrow f}}
& = &
\frac{\lambda_{M \rightarrow f} + \xi_{P^0 \rightarrow M}}{
1 + \xi_{P^0 \rightarrow M} \lambda_{M \rightarrow f}}\ ,
\\*[4mm]
\chi_2 & \equiv &
\frac{q_P}{p_P}
\frac{A_{\overline{P^0} \rightarrow \overline{M^0}} \frac{p_M}{q_M}
A_{M^0 \rightarrow f} +
A_{\overline{P^0} \rightarrow M^0} \frac{q_M}{p_M}
A_{\overline{M^0} \rightarrow f}}{
A_{P^0 \rightarrow M^0} A_{M^0 \rightarrow f} +
A_{P^0 \rightarrow \overline{M^0}} A_{\overline{M^0} \rightarrow f}}
& = &
\lambda_{P \rightarrow M^0}
\frac{\xi_{\overline{P^0} \rightarrow M} + \lambda_{M \rightarrow f}}{
1 + \xi_{P^0 \rightarrow M} \lambda_{M \rightarrow f}}\ ,
\\*[4mm]
\chi_3 & \equiv &
\frac{q_P}{p_P}
\frac{A_{\overline{P^0} \rightarrow \overline{M^0}}
A_{\overline{M^0} \rightarrow f} +
A_{\overline{P^0} \rightarrow M^0}
A_{M^0 \rightarrow f}}{
A_{P^0 \rightarrow M^0} A_{M^0 \rightarrow f} +
A_{P^0 \rightarrow \overline{M^0}} A_{\overline{M^0} \rightarrow f}}
&=&
\lambda_{P \rightarrow M^0}
\frac{\xi_{\overline{P^0} \rightarrow M} \lambda_{M \rightarrow f} + 1}{
1 + \xi_{P^0 \rightarrow M} \lambda_{M \rightarrow f}}\ .
\end{array}
\end{equation}
Notice that the parameters $\chi_n$ involve
$\xi_{P^0 \rightarrow M}$ and $\xi_{\overline{P^0} \rightarrow M}$,
which describe the new type of interference effects which are
the subject of this article.

However,
$\chi_3$ may always be reinterpreted in a different way.
Let us consider the state
\begin{equation}
\label{state into f}
| M_{{\rm into\, } f} \rangle
\equiv
A^\ast_{M^0 \rightarrow f} | M^0 \rangle
+
A^\ast_{\overline{M^0} \rightarrow f} | \overline{M^0} \rangle\ .
\end{equation}
This is the $M^0 - \overline{M^0}$ combination that
decays into the final state $f$ at time $t_M$.
Indeed, this state is orthogonal to the state
\begin{equation}
| M_{{\rm not\,  into\, } f} \rangle
\equiv
A_{\overline{M^0} \rightarrow f} | M^0 \rangle
-
A_{M^0 \rightarrow f} | \overline{M^0} \rangle\ ,
\end{equation}
which clearly cannot decay into the final state $f$ because
$\langle f | T | M_{{\rm not\,  into\, } f} \rangle = 0$.
Using Eq.~\ref{state into f},
it is easy to show that
\begin{equation}
\label{chi3 e lambda}
\chi_3 = \lambda_{P \rightarrow M_{{\rm into\, } f}}.
\end{equation}
Still,
in general,
$\chi_1$ and $\chi_2$ may not be reinterpreted as parameters $\lambda$.

One may wonder why these effects did not show up in the
analysis presented in \cite{Azimov,Dass,Kay96,Kay97,AziDun}.
The reason is simple:
in those cases a meson $P^0$ can only decay into one of the 
flavour eigenstates of the intermediate meson system.
For example,
$P^0$ can decay into $M^0$, but not into $\overline{M^0}$.
Indeed,
taking 
$A_{\overline{P^0} \rightarrow M^0} = 0 = A_{P^0 \rightarrow \overline{M^0}}$,
we find,
in addition to Eq.~\ref{chi3 e lambda},
\begin{equation}
\label{before}
\chi_1 = \lambda_{M \rightarrow f}\ \ \mbox{and }\ \ 
\chi_2 = \lambda_{P \rightarrow M_H}.
\end{equation}
The last equality involves $M_H$,
the heavy mass eigenstate of the $M^0 - \overline{M^0}$ system,
and holds if and only if $|p_M/q_M|=1$.
In these cases,
we do not need to introduce the parameters $\xi$.
On the contrary,
these parameters are required in the discussion of the
$B^+ \rightarrow D K^+ \rightarrow [f]_D K^+$ cascade decays,
as found by Meca and Silva \cite{Silva}.
The same parameters would enter the analysis of the decays,
$B^+ \rightarrow D_{H,L} K^+$,
if only these decays could be disentangled from one another \cite{Xing}.
Unfortunately, they cannot \cite{Xing}.
It should now be clear that the effects described by 
$\xi_{P^0 \rightarrow M}$
and $\xi_{\overline{P^0} \rightarrow M}$ are mandatory and,
when combined with the usual CP-violating observables,
are also sufficient for a discussion of the $B \rightarrow D$
and $D \rightarrow K$ decay chains.

In the example just discussed,
in which
$A_{\overline{P^0} \rightarrow M^0} = 0 = A_{P^0 \rightarrow \overline{M^0}}$,
we find that $\chi_3 = \chi_1 \chi_2$.
This relation, although not valid in general,
holds in a variety of special cases.
Indeed,
\begin{equation}
\chi_3 - \chi_1 \chi_2
=
\frac{\left( 1 - \lambda^2_{M \rightarrow f}\right) 
\left( 1 - \xi_{\overline{P^0} \rightarrow M} \xi_{P^0 \rightarrow M} \right)
\lambda_{P \rightarrow M^0}}{
\left(
1 + \xi_{P^0 \rightarrow M} \lambda_{M \rightarrow f}
\right)^2}.
\end{equation}
This result is completely general.
We conclude that $\chi_3 - \chi_1 \chi_2$ vanishes
if any of the following conditions holds:
\begin{enumerate}
\item $\lambda^2_{M \rightarrow f} = 1$, which, if $f$ is a CP eigenstate,
means that there is CP conservation when the
$M^0 - \overline{M^0}$ system decays into the final state $f$;
\item $\xi_{\overline{P^0} \rightarrow M} \xi_{P^0 \rightarrow M} = 1$,
meaning that there is CP conservation when the
$P^0 - \overline{P^0}$ system decays into the $M^0 - \overline{M^0}$ system;
\item $\lambda_{P \rightarrow M^0}=0$, meaning that the decay
$\overline{P^0} \rightarrow M^0 X$ is forbidden.
Analogously, the difference also vanishes if
$P^0 \rightarrow M^0 X$ is forbidden.
\end{enumerate}
If any of these conditions holds,
the amplitude for the cascade decay depends only on two independent
parameters (for example $\chi_1$ and $\chi_2$),
in addition to the overall normalization factor.
We see that $\chi_3 - \chi_1 \chi_2$ is only
different from zero if there are interference effects in every step of
the decay chain.\footnote{Informally,
we may think of $\chi_3 - \chi_1 \chi_2$ as the ``all hell breaks loose''
parameter.}

We now turn to the CP invariance constraints on the $\chi_n$ parameters.
We find that,
if there is CP conservation, then
\begin{equation}
\label{CPcons chi2}
\chi_2 = \eta_X \eta_P \eta_M,
\end{equation}
and $\chi_3 = \chi_1 \chi_2$.
In addition,
if $f$ is a CP eigenstate,
then we find two further conditions for CP invariance:
\begin{equation}
\chi_1 = \eta_M \eta_f\ \ \mbox{and }\ \ 
\chi_3 = \eta_X \eta_P \eta_f.
\end{equation}

We now consider the case in which we start with a tagged
$\overline{P^0}$. The decay amplitude for the cascade chain becomes
\begin{eqnarray}
\label{amplitude:Pbar}
A \left[ \overline{P^0} \stackrel{t_P}{\rightarrow} M 
\stackrel{t_M}{\rightarrow} f \right]
&=&
\frac{p_P}{q_P}\, 
\left( 
A_{P^0 \rightarrow M^0} A_{M^0 \rightarrow f} + 
A_{P^0 \rightarrow \overline{M^0}} A_{\overline{M^0} \rightarrow f}
\right) \times
\nonumber\\
& &
\hspace{2mm}
\left[
\chi_3\, g_+^P g_+^M +
\chi_2\,  g_+^P g_-^M +
\chi_1\, g_-^P g_-^M +
g_-^P g_+^M
\right]\ .
\end{eqnarray}
This result may be obtained from Eq.~\ref{amplitude:P} by
interchanging $g_+^P \leftrightarrow g_-^P$ and multiplying the
result by $p_P/q_P$.

\subsection{Decay rates}

Using Eq.~\ref{amplitude:P}, 
we find for the decay
$P^0 \rightarrow  M X \rightarrow [f]_M X$,
\begin{eqnarray}
\label{rate from P0}
\Gamma \left[ P^0 \stackrel{t_P}{\rightarrow} M 
\stackrel{t_M}{\rightarrow} f \right]
&=&
N \left[
\left| g_+^P g_+^M \right|^2
+ |\chi_1|^2 \left| g_+^P g_-^M \right|^2
+ |\chi_2|^2 \left| g_-^P g_-^M \right|^2
+ |\chi_3|^2 \left| g_-^P g_+^M \right|^2
\right.
\nonumber\\
&+&
2 \left| g_+^P \right|^2
\mbox{Re } \left\{ \chi_1 g_+^{M \ast} g_-^M \right\}
+
2 \left| g_-^P \right|^2
\mbox{Re } \left\{ \chi_2 \chi_3^\ast  g_+^{M \ast} g_-^M \right\}
\nonumber\\
&+&
2 \left| g_+^M \right|^2
\mbox{Re } \left\{ \chi_3 g_+^{P \ast} g_-^P \right\}
+
2 \left| g_-^M \right|^2
\mbox{Re } \left\{ \chi_1^\ast \chi_2 g_+^{P \ast} g_-^P \right\}
\nonumber\\
&+&
\left.
2 \mbox{Re } \left\{ \chi_2 g_+^{P \ast} g_-^P g_+^{M \ast} g_-^M  \right\}
+
2 \mbox{Re } \left\{ \chi_1^\ast \chi_3 g_+^{P \ast} g_-^P 
g_+^M g_-^{M \ast}  \right\}
\right]\ ,
\end{eqnarray}
where
\begin{equation}
N \equiv
\left|
A_{P^0 \rightarrow M^0} A_{M^0 \rightarrow f} + 
A_{P^0 \rightarrow \overline{M^0}} A_{\overline{M^0} \rightarrow f}
\right|^2\ .
\end{equation}
Similarly,
the decay rate for an initial $\overline{P^0}$ is
\begin{eqnarray}
\label{rate from P0bar}
\frac{\Gamma \left[ \overline{P^0} \stackrel{t_P}{\rightarrow} M 
\stackrel{t_M}{\rightarrow} f \right]}{|p_P/q_P|^2}
&=&
N
\left[
|\chi_3|^2 \left| g_+^P g_+^M \right|^2
+ |\chi_2|^2 \left| g_+^P g_-^M \right|^2
+ |\chi_1|^2 \left| g_-^P g_-^M \right|^2
+ \left| g_-^P g_+^M \right|^2
\right.
\nonumber\\
&+&
2 \left| g_+^P \right|^2
\mbox{Re } \left\{ \chi_2 \chi_3^\ast  g_+^{M \ast} g_-^M \right\}
+
2 \left| g_-^P \right|^2
\mbox{Re } \left\{ \chi_1 g_+^{M \ast} g_-^M \right\}
\nonumber\\
&+&
2 \left| g_+^M \right|^2
\mbox{Re } \left\{ \chi_3^\ast g_+^{P \ast} g_-^P \right\}
+
2 \left| g_-^M \right|^2
\mbox{Re } \left\{ \chi_1 \chi_2^\ast g_+^{P \ast} g_-^P \right\}
\nonumber\\
&+&
\left.
2 \mbox{Re } \left\{ \chi_1 \chi_3^\ast g_+^{P \ast} g_-^P 
g_+^{M \ast} g_-^M  \right\}
+
2 \mbox{Re } \left\{ \chi_2^\ast g_+^{P \ast} g_-^P
g_+^M g_-^{M \ast} \right\}
\right]\ ,
\end{eqnarray}

One may also use the rates in Eqs.~\ref{rate from P0} and
\ref{rate from P0bar} to describe decays of the type
$P^\pm \rightarrow  M X^\pm \rightarrow [f]_M X^\pm$.
To this end,
one substitutes $P^0 \rightarrow P^+$,
$\overline{P^0} \rightarrow P^-$,
$g^P_+ \rightarrow \exp{(- \Gamma_{P^+} t_P / 2)}$,
$g^P_- \rightarrow 0$,
and $q_P/p_P \rightarrow 1$.

\section{Classification of cascade decays}

To simplify the notation we will suppress the explicit reference to the
final stage of the decay chain, $[f]_M X$,
unless it is required.
Traditionally,
one has considered cascade decays of the type
$B_d, B_s \rightarrow K$,
using the known mixing parameters of the $K^0 - \overline{K^0}$
system as an analyzer for the parameters of the heavier
$B_d^0 - \overline{B_d^0}$ \cite{Azimov,Dass,Kay96,Kay97}
and $B_s^0 - \overline{B_s^0}$ \cite{AziDun} systems.
More recently,
Meca and Silva \cite{Silva} 
have stressed the role of $B^\pm \rightarrow D$ decays as a 
probe for new physics contributions to the $D^0 - \overline{D^0}$
mixing. 

Clearly,
cascade decays of the type
$P^\pm \rightarrow  M X^\pm \rightarrow [f]_M X^\pm$
are useful for the second strategy,
while cascade decays of the type
$P \rightarrow  M X \rightarrow [f]_M X$ could,
in principle,
be useful for both.
We analyze them in turn.

\subsection{Cascade decays from charged $P$ mesons}

Decays of this type include
$D^+ \rightarrow \overline{K^0} K^+$,
$D^+ \rightarrow \{ \overline{K^0}, K^0 \} \pi^+$,
$D_s^+ \rightarrow K^0 \pi^+$,
and $D_s^+ \rightarrow \{ K^0, \overline{K^0} \} K^+$.
Since the kaon system is well tested experimentally,
these decays are of limited use.
This is also true of the pure penguin decays
$B^+ \rightarrow K^0 \pi^+$
and $B^+ \rightarrow \overline{K^0} K^+$.

More interesting are the decays
$B^+ \rightarrow \{ D^0, \overline{D^0} \} K^+$,
such as the one represented in Fig.~2.
\begin{figure}
\centerline{\psfig{figure=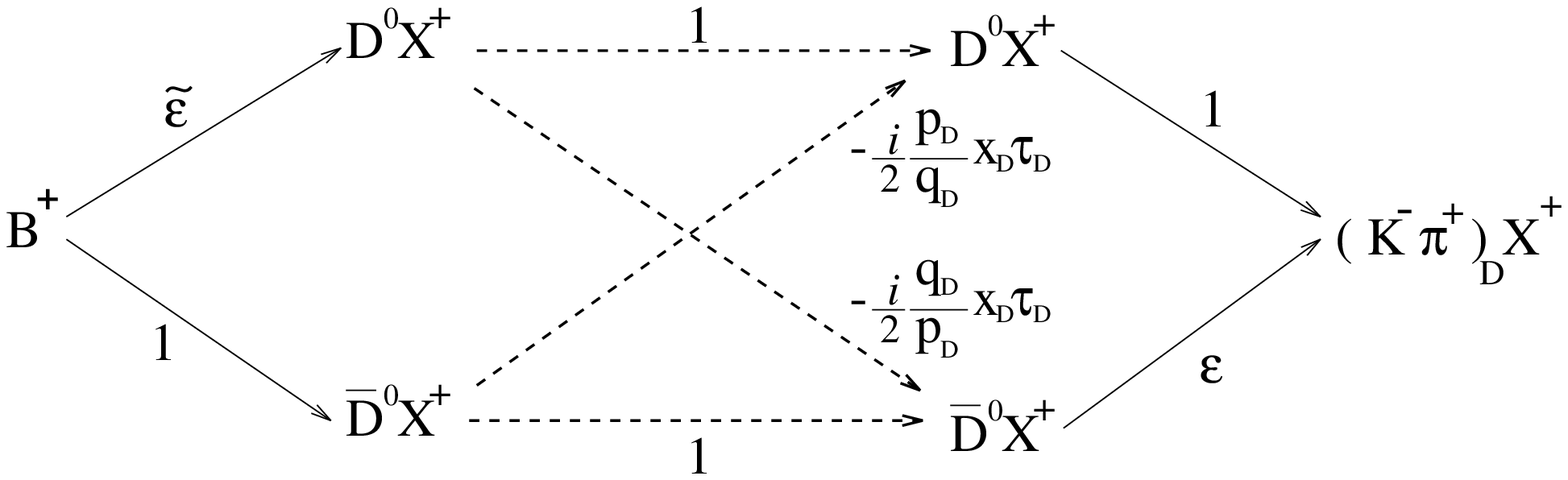,height=1.5in}}
\caption{The $B^+ \rightarrow D X^+ \rightarrow (K^- \pi^+)_D X^+$
cascade decay.
\label{fig:2}}
\end{figure}
We use
\begin{equation}
\epsilon \equiv \left|
\frac{A_{\overline{D^0} \rightarrow K^- \pi^+}}{
A_{D^0 \rightarrow K^- \pi^+}}
\right|
\sim
\left|
\frac{V_{cd} V_{us}}{V_{cs} V_{ud}}
\right|
\sim 0.05\ ,
\end{equation}
and \cite{ADS}
\begin{equation}
\tilde{\epsilon} \equiv \left|
\frac{A_{B^+ \rightarrow D^0 K^+}}{
A_{B^+ \rightarrow \overline{D^0} K^+}}
\right|
\sim
\left|
\frac{V_{ub} V_{cs}}{V_{cb} V_{us}}
\frac{a_2}{a_1}
\right|
\sim 0.09\ ,
\end{equation}
where $|a_2/a_1| \sim 0.26$ accounts for the fact that
the $B^+ \rightarrow D^0 K^+$ decay is color-suppressed,
while the $B^+ \rightarrow \overline{D^0} K^+$ decay is not.

Meca and Silva have pointed out that these decays may be used to get 
at $x_D \equiv \Delta m_D/\Gamma_D$ \cite{Silva}.
This is easy to see from Fig.~2,
where we have taken $y_D=0$,
and used the small $x_D$ approximation,
$g_+^D \sim 1$ and $g_-^D \sim -i/2\, x_D \tau_D$,
with $\tau_D \equiv \Gamma_D t_D$.
If $\Delta m_D = 0$,
then there are no mixed decay paths and we only have two contributions
to the decay amplitude.
In suitable units,
one is of order $\epsilon$ and the other of order $\tilde{\epsilon}$.
On the other hand,
if $\Delta m_D \neq 0$,
then there are two further decay paths;
one is of order $x_D$ and the other of order
$\epsilon \tilde{\epsilon} x_D$.
Now,
values of $x_D \sim 10^{-2}$ are easy to obtain
in many models of new physics
\cite{several}.
In that case,
the last term may be dropped,
but the term linear in $x_D$ corrects the unmixed
$\epsilon$ and $\tilde{\epsilon}$ terms by about $10\%$.
This has a corresponding impact on the
Gronau-London-Wyler \cite{GL,GW} and Atwood-Dunietz-Soni \cite{ADS}
methods to get at the CKM phase $\gamma$ with the
$B^+ \rightarrow \{ D^0, \overline{D^0} \} K^+$ decays \cite{Silva}.

The decay rates for chains like the one in Fig.~2 are easy to calculate.
Using Eq.~\ref{rate from P0} we find,
\begin{eqnarray}
\label{rate for B+ -> D}
\Gamma \left[ B^+ \stackrel{t_B}{\rightarrow} D
\stackrel{t_D}{\rightarrow} f \right]
&=&
e^{- \Gamma_{B^+} t_B}
\left|
A_{B^+ \rightarrow D^0} A_{D^0 \rightarrow f} +
A_{B^+ \rightarrow \overline{D^0}} A_{\overline{D^0} \rightarrow f}
\right|^2
\times
\nonumber\\
& &
\left[
\left| g_+^D \right|^2
+ 2 \mbox{Re}\, \left\{ \chi_1 g_+^{D \ast} g_-^D \right\}
+ |\chi_1|^2 \left| g_-^D \right|^2
\right]\ ,
\end{eqnarray}
where
\begin{equation}
\chi_1 \equiv \frac{\lambda_{D \rightarrow f} + \xi_{B^+ \rightarrow D}}{
1 + \lambda_{D \rightarrow f}\, \xi_{B^+ \rightarrow D}}.
\end{equation}
Taking $y_D=0$ and the small $x_D$ approximation,
the term dependent on $g_+^{D \ast} g_-^D$ is proportional to
$\mbox{Im} \chi_1\, x_D$.
It contributes to the rate due to two independent interference effects:
one related to $\mbox{Im} \lambda_{D \rightarrow f}$,
the other related to $\mbox{Im} \xi_{B^+ \rightarrow D}$
(see Eq.~\ref{not so obvious} bellow).
This is easiest to understand with the aid of the
parametrizations
\begin{eqnarray}
A_{D^0 \rightarrow K^- \pi^+} = A\ ,
& \hspace{5mm} &
A_{\overline{D^0} \rightarrow K^- \pi^+} =
- \epsilon \, A\, e^{i \Delta_D} ,
\nonumber\\
A_{B^+ \rightarrow \overline{D^0} K^+}
=
B\ ,
& \hspace{5mm} &
A_{B^+ \rightarrow D^0 K^+}
=
\tilde{\epsilon}\, e^{i\gamma}\, B\, e^{i \Delta_B} ,
\end{eqnarray}
where $A$, and $B$ contain form factors, strong phases and magnitudes
of CKM matrix elements.
$\Delta_D$ and $\Delta_B$ are the differences of strong phases that exist
between the amplitudes involving $D^0$ and the corresponding amplitudes
involving $\overline{D^0}$.
In addition,
we take $q_D/p_D = e^{2 i \theta_D}$,
thus allowing for the fact that the new physics effects might also
bring with them new phases into the $D^0 - \overline{D^0}$ mixing.
Therefore,
\begin{equation}
\lambda_{D \rightarrow K^- \pi^+} =
- \epsilon\, e^{i(2 \theta_D + \Delta_D)}
\ \ \ \mbox{and }\ \ \ 
\xi_{B^+ \rightarrow D K^+} =
\frac{1}{\tilde{\epsilon}}\, e^{-i(\gamma + 2 \theta_D + \Delta_B)}\ .
\end{equation}
The term linear in $x_D$ becomes proportional to
\begin{eqnarray}
\label{not so obvious}
\mbox{Im}\, \chi_1
& \propto &
\mbox{Im}\, \lambda_{D \rightarrow K^- \pi^+}
\left( 1 - |\xi_{B^+ \rightarrow D K^+}|^2 \right)
+ \mbox{Im}\, \xi_{B^+ \rightarrow D K^+}
\left( 1 - |\lambda_{D \rightarrow K^- \pi^+}|^2 \right)
\nonumber\\
& \sim &
\frac{1}{\tilde{\epsilon}^2}
\left[
\epsilon \sin{(2 \theta_D + \Delta_D)}
- \tilde{\epsilon} \sin{(\gamma + 2 \theta_D + \Delta_B)}
\right]\ ,
\end{eqnarray}
where we have neglected higher order terms in $\epsilon$ and
$\tilde{\epsilon}$.
On the one hand,
there is a contribution proportional to $\epsilon x_D$ which arises from
the $\mbox{Im} \lambda_{D \rightarrow f}$ piece in $\chi_1$.
This involves $\theta_D$, which vanishes in the SM,
and the difference of strong phases $\Delta_D$,
which is difficult to estimate but could also be small.
Obviously,
these two effects are the same that appear in the direct decay
$D \rightarrow K^- \pi^+$;
they were studied in this context by \cite{Wol95},
and \cite{Bla95,Bro95},
respectively.
On the other hand,
there is a contribution proportional to $\tilde{\epsilon} x_D$ which
arises from the $\mbox{Im} \xi_{B^+ \rightarrow D}$ piece in $\chi_1$.
This is {\em guaranteed} large even in the SM,
for it involves the large CKM phase $\gamma$.
Therefore,
these decays may be used to probe values of $x_D \sim 10^{-2}$.
This case is discussed in detail in reference \cite{Silva}.

Decays like $B^+ \rightarrow \{ D^0, \overline{D^0} \} \pi^+$ 
and $B^+ \rightarrow \{ D^0, \overline{D^0} \} \rho^+$
might also be useful in the search for $x_D$. 
These decay chains have much larger branching ratios than
$B^+ \rightarrow \{ D^0, \overline{D^0} \} K^+$.
Indeed,
$\mbox{BR}[B^+ \rightarrow \overline{D^0} \rho^+]
= (1.34 \pm 0.18) \times 10^{-2}$,
$\mbox{BR}[B^+ \rightarrow \overline{D^0} \pi^+]
= (5.3 \pm 0.5) \times 10^{-3}$ \cite{PDG},
while
$\mbox{BR}[B^+ \rightarrow \overline{D^0} K^+]/
\mbox{BR}[B^+ \rightarrow \overline{D^0} \pi^+]
= 0.055 \pm 0.015$ \cite{measure}.
This means that the overall normalization factor $|B|^2$
in Eq.~\ref{rate for B+ -> D} is roughly $18$ ($46$) times larger in
$B^+ \rightarrow D \pi^+$
($B^+ \rightarrow D \rho^+$)
decays than it is in $B^+ \rightarrow D K^+$ decays.
But,
in these cases,
$\tilde{\epsilon}$ is further suppressed by a factor of
about $0.05$ with respect to its value in the
$B^+ \rightarrow \overline{D^0} K^+$ decay chains.
This suppresses the $\epsilon \tilde{\epsilon}$ and 
$\tilde{\epsilon} x_D$ interference terms in the decay rate,
meaning that these interference effects come into all the decay
chains at roughly the same level.
On the contrary,
the term proportional to $\epsilon x_D$ is the same for all the decay chains.
Therefore,
it should be much easier to detect in $B^+ \rightarrow D \pi^+$
and $B^+ \rightarrow D \rho^+$
decays than it is in the original $B^+ \rightarrow D K^+$ decays
proposed by Meca and Silva \cite{Silva}.
We recall that the $\epsilon x_D$ term involves
$\mbox{Im} \lambda_{D \rightarrow f}$ which also shows up in direct
$D \rightarrow f$ decays.
Hence,
it is only detectable if the new physics also produces large CP-violating
effects, $\theta_D$, in the $D^0 - \overline{D^0}$ system \cite{Wol95},
or if the strong phase difference, $\Delta_D$, is considerable
\cite{Bla95,Bro95}.

The crucial role played by the $\xi_i$ parameters in the 
$B^+ \rightarrow \{ D^0, \overline{D^0} \} K^+ \rightarrow [f]_D K^+$
cascade decays is seen most vividly by taking $f = X^- l^+ \nu_l$.
This case is illustrated in Fig.~3.
\begin{figure}
\centerline{\psfig{figure=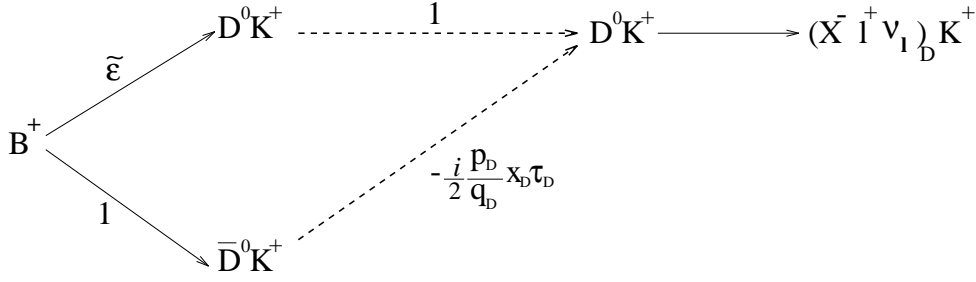,height=1.5in}}
\caption{The $B^+ \rightarrow D K^+ \rightarrow
(X^- l^+ \nu_l)_D K^+$ cascade decay. 
\label{fig:3}}
\end{figure}
Since the decay $\overline{D^0} \rightarrow X^- l^+ \nu_l$ is forbidden,
we have $\lambda_{D \rightarrow f} = 0$,
$\chi_1 = \xi_{B^+ \rightarrow D}$
and, thus,
the interference is determined solely by $\xi_{B^+ \rightarrow D}$.
Indeed,
it is clear from Fig.~3 that the only interference effect present
is that arising from the clash between the $D^0 - \overline{D^0}$ mixing
and the decay from the $B^+$ {\em into that system}.
Notice the crucial role played by the fact that there are two decay
paths from $B^+$ into the $D^0 - \overline{D^0}$ system;
$B^+ \rightarrow D^0$
and $B^+ \rightarrow \overline{D^0}$.
Both must be present in order
for the decay chain to display this new type of interference.

Finally,
there are $B^+ \rightarrow \overline{D^0} D_s^+$ and
$B^+ \rightarrow \overline{D^0} D^+$ decays.
However,
there are no $B^+ \rightarrow D^0 D_s^+$ and
$B^+ \rightarrow D^0 D^+$ decays.
As a result, 
in Fig.~2 there is only one decay path from the initial $B^+$
into the intermediate state and,
thus,
the effects of that decay factor out in the overall decay rates.
Everything works as if we had started from an initial
$\overline{D^0}$ meson.

\subsection{Cascade decays $B \rightarrow K$}

Let us consider decay chains of the type
$B^0_d \rightarrow K^0 X \rightarrow [f]_K X$,
where $X$ may be $J/\psi$, $\phi$, $\omega$, $\rho^0$,
$\pi^0$, etc.
In this case,
we have $A_{B_d^0 \rightarrow \overline{K^0}} = 0$
and $A_{\overline{B_d^0} \rightarrow K^0} = 0$ in the SM.
Therefore,
the parameters needed for Eqs.~\ref{rate from P0} and
\ref{rate from P0bar} become
\begin{eqnarray}
\label{no caso 1}
N & = & \left| A_{B^0_d \rightarrow K^0} A_{K^0 \rightarrow f} \right|^2\ ,
\nonumber\\
\chi_1 & = & \lambda_{K \rightarrow f}\ ,
\nonumber\\
\chi_2 & = & \frac{q_{B_d}}{p_{B_d}} \frac{p_K}{q_K}
\frac{A_{\overline{B^0_d} \rightarrow \overline{K^0}}}{
A_{B^0_d \rightarrow K^0}}
= - \lambda_{B_d \rightarrow K_S}\ ,
\nonumber\\
\chi_3 & = & \chi_1 \chi_2\ .
\end{eqnarray}
We had already anticipated these results in Eq.~\ref{before},
where we mentioned that the last equality on the third line holds if 
$|q_K/p_K|=1$.
From the experimental measurement of CP violation in 
the $K^0 - \overline{K^0}$ system,
we know that this equality holds to the level of $10^{-3}$.

The situation is very similar for the decay chains of the type
$B^0_s \rightarrow \overline{K^0} X \rightarrow [f]_K X$.
Here,
$A_{B_s^0 \rightarrow K^0} = 0$
and $A_{\overline{B_s^0} \rightarrow \overline{K^0}} = 0$ in the SM.
Hence,
\begin{eqnarray}
\label{no caso 2}
N & = & \left| A_{B^0_s \rightarrow \overline{K^0}}
A_{\overline{K^0} \rightarrow f} \right|^2\ ,
\nonumber\\
\chi_1 & = & 1 / \lambda_{K \rightarrow f}\ ,
\nonumber\\
\chi_2 & = & \frac{q_{B_s}}{p_{B_s}} \frac{q_K}{p_K}
\frac{A_{\overline{B^0_s} \rightarrow K^0}}{
A_{B^0_s \rightarrow \overline{K^0}}}
= - \lambda_{B_s \rightarrow K_S}\ ,
\nonumber\\
\chi_3 & = & \chi_1 \chi_2\ .
\end{eqnarray}
The articles \cite{Azimov,Dass,Kay96,Kay97}
discuss the first case, with $X = J/\psi$;
while reference \cite{AziDun} describes the second case,
also for $X = J/\psi$.
As we have stressed before,
in these cases,
the fact that there is only one decay path from the $B^0$ into the
$K^0 - \overline{K^0}$ system implies that all the interference effects
(and any CP violation therein)
may be written in terms of the classic $\lambda$ parameters.

To find the decay rates,
we could substitute Eq.~\ref{no caso 1} or Eq.~\ref{no caso 2}
into Eq.~\ref{rate from P0}.
However,
we know that the lifetimes of the mass eigenstates of the
$K^0 - \overline{K^0}$ system are very different.
Therefore,
it is more appropriate to write our expressions in terms of $\gamma_S$
and $\gamma_L$,
which are the lifetimes of the short-lived ($K_S$)
and long-lived ($K_L$) mass eigenstates, respectively.
Since in both cases $\chi_3 = \chi_1 \chi_2$,
Eq.~\ref{amplitude:P} may be rewritten as
\begin{equation}
\label{amplitude: B -> K}
A \left[ B^0 \stackrel{t_B}{\rightarrow} K 
\stackrel{t_K}{\rightarrow} f \right]
\propto
e^{-i \mu_H^K t_K} \left( 1 + \chi_1 \right)
\left( g_+^B + \chi_2 g_-^B \right)
+
e^{-i \mu_L^K t_K} \left( 1 - \chi_1 \right)
\left( g_+^B - \chi_2 g_-^B \right)\ .
\end{equation}
Defining
\begin{equation}
e^{i \phi} \equiv
- \frac{(1 + \chi_1)(1 - \chi_1^\ast)}{|(1 + \chi_1)(1 - \chi_1^\ast)|},
\end{equation}
the decay rate becomes
\begin{eqnarray}
\label{AziKay}
& &
4\, \Gamma \left[ B^0 \stackrel{t_B}{\rightarrow} K
\stackrel{t_K}{\rightarrow} f \right]\, / N =
\nonumber\\
& &
\hspace{15mm}
e^{- \gamma_S t_K} \left| 1 - \chi_1 \right|^2
\left| g_+^B - \chi_2 g_-^B \right|^2
+ e^{- \gamma_L t_K} \left| 1 + \chi_1 \right|^2
\left| g_+^B + \chi_2 g_-^B \right|^2
\nonumber\\
& &
\hspace{15mm}
- 2 e^{- \frac{\gamma_S + \gamma_L}{2} t_K}
\left| 1 - \chi_1 \right| \left| 1 + \chi_1 \right|
\left[ \cos{(\Delta m_K\, t_K - \phi)}
\left( \left| g_+^B \right|^2 - \left| \chi_2 g_-^B \right|^2 \right)
\right.
\nonumber\\
& &
\hspace{55mm}
\left.
+ 2 \sin{(\Delta m_K\, t_K - \phi)}
\mbox{Im} \left( \chi_2 g_+^{B \ast} g_-^B \right)
\right].
\end{eqnarray}
Now, using Eqs.~\ref{eigenvectors} and recalling that $\chi_1$
is either equal to $\lambda_f$ or to $1/\lambda_f$,
we notice that
\begin{eqnarray}
|1 - \chi_1| & \propto & \left| A_{K_S \rightarrow f} \right|\ ,
\nonumber\\
|1 + \chi_1| & \propto & \left| A_{K_L \rightarrow f} \right|\ .
\end{eqnarray}
Therefore,
the first term in Eq.~\ref{AziKay} describes the decays in which the
final state $f$ has come from $K_S$.
Taking $f = \pi^+ \pi^-$,
this is the dominant term for times $t_K < 1/\gamma_S$.
Likewise,
the second term refers to the decays in which the final state $f$
has come from $K_L$.
At late times,
$t_K \gg 1/\gamma_S$,
the cascade chain is dominated by these decays since
by then the $K_S$ component will have decayed away.
The last terms describe the interference between the
path going through $K_S$ and that going through $K_L$.
These will become important at intermediate times \cite{Azimov,Kay97}.

Let us consider decays of the type $B_d \rightarrow K J/\psi$.
In the SM,
these decays get tree level and penguin contributions which,
to an excellent approximation,
share a common weak phase.
Moreover,
in the SM one has $\Delta \Gamma_{B_d} \sim 0$ and $|q_B/p_B| \sim 1$.
As a result $|\chi_2|^2=1$,
$\mbox{Re} \chi_2 = - \cos 2 \beta$,
and $\mbox{Im} \chi_2 = \sin 2 \beta$.
Also,
since $\Delta \Gamma_B \sim 0$,
we get from Eq.~\ref{g+*g-2} that $\mbox{Re} (g_+^{B \ast} g_-^B) \sim 0$.
This is the origin of the standard 
result\footnote{It
should be stressed
that the classic $B_d \rightarrow J/\psi K_S$ experiments involve,
in fact, cascade decays,
since the kaon will ultimately be detected through its decay into
$\pi^+ \pi^-$.
However,
since the $K_L$ is mostly CP odd,
the detected $\pi^+ \pi^-$ is very unlikely to have come from a $K_L$
($\chi_1 \sim \eta_K = - 1$ in Eqs.~\ref{eq:CP2} and \ref{AziKay}).
Moreover,
the experiments will select events with low $t_K$ \cite{Kayprivate}.
We see from Eq.~\ref{AziKay} that,
under these circumstances,
the rate is overwhelmingly dominated by the first term,
which corresponds to the pure $B_d \rightarrow J/\psi K_S$
decay \cite{Kayprivate}.
}:
the pure $K_S$ (and the pure $K_L$) term of Eq.~\ref{AziKay}
only measures $\sin 2 \beta$,
leaving a fourfold ambiguity in the determination of $\beta$.
Kayser has pointed out that one can use the $K_S - K_L$ interference
on the last line of Eq.~\ref{AziKay} to get at $\cos 2 \beta$,
thus reducing the ambiguity \cite{Kay97}.

One may also consider those decay chains where the primary decay
is $B^0_d \rightarrow K^0 \rho^0$,
$B^0_d \rightarrow K^0 \pi^0$, etc.
If these primary decays are dominated by the top-mediated penguin diagram,
then $\chi_2$ measures again the weak phase $\beta$.
However,
although suppressed by $|V_{ub} V_{us}/(V_{tb} V_{ts})|$ and by color,
the tree level diagram comes in with a different weak phase.
The resulting hadronic uncertainties imply that these decay chains
cannot compete with $B_d \rightarrow K J/\psi$ in the determination
of the CKM phase $\beta$.

We now turn to decays of the type
$B^0_s \rightarrow \overline{K^0} J/\psi$.
The considerations made about the
$B_d \rightarrow K J/\psi$ case also
apply here,
with one exception.
Here, the SM tree level and penguin contributions have a
large weak phase difference,
which turns out to be equal to $\beta$.
In fact,
the decay amplitudes may be parametrized as\footnote{In these expressions,
we have used the standard phase convention for the CKM matrix elements,
and we have ignored the CP transformation phases of $B_s$ and $K$.
However,
to be consistent,
we have kept the $\eta_X = -1$ factor obtained
in getting from the first to the second line \cite{Thank you}.}
\begin{eqnarray}
\label{Eq do AziDun}
A_{B^0_s \rightarrow \overline{K^0} J/\psi}
=
V_{cb}^\ast V_{cd} T + V_{tb}^\ast V_{td} P
\sim
- \lambda^3 T + \lambda^3 e^{- i \beta} P,
\nonumber\\
- A_{\overline{B^0_s} \rightarrow K^0 J/\psi}
=
V_{cb} V_{cd}^\ast T + V_{tb} V_{td}^\ast P
\sim
- \lambda^3 T + \lambda^3 e^{+ i \beta} P.
\end{eqnarray}
The symbols $T$ and $P$ stand for the tree level and penguin diagram
contributions, with the CKM factor taken out;
they include form-factors,
strong phases,
the gauge couplings (weak and strong, respectively),
and $P$ also includes the usual one-loop prefactor.
Since the penguin diagram is not CKM suppressed with respect to
the tree level diagram,
$|\chi_2| = |\lambda_{B_s \rightarrow J/\psi K_S}|$ may differ from one
by a few percent.
Azimov and Dunietz have pointed out that one may use these decay chains
to get at the elusive $\Delta m_{B_s}$ mass difference \cite{AziDun}. 
It is well known that $\Delta m_{B_s}$ is difficult to determine
from time-dependent measurements because of the vertexing limits.
These preclude measurements of $x_s$ much larger than about $20$.
The situation is also bleak if we integrate the decays over $t_B$
because the effects of $x_s$ show up in the rate into a flavour-specific
final state as $1/(1 + x^2)$.
This should be compared to what one gets by integrating Eq.~\ref{AziKay}
over $t_B$.
As shown in the appendix,
the result may be gotten in a straightforward way with the substitutions
$|g_+^B|^2 \rightarrow G_+^B$,
$|g_-^B|^2 \rightarrow G_-^B$,
and $g_+^{B \ast} g_-^B \rightarrow G_{+-}^B$.
Now,
Azimov and Dunietz point out that,
upon integration over $t_B$,
the last term of Eq.~\ref{AziKay} contains a term proportional
to $\mbox{Re} \chi_2\, \mbox{Im} G^B_{+-}$.
And, the function $\mbox{Im} G^B_{+-}$ shown in Eq.~\ref{G+-} involves
$x/(1 + x^2)$ rather than $1/(1 + x^2)$ \cite{AziDun}.
This increases the sensitivity to $x_s$ considerably.
Moreover,
this term involves $\mbox{Re} \chi_2$ which is nonzero even in
the limit of CP conservation \cite{AziDun}.
To appreciate this effect,
we take $x_s \gg 1$, $|y_s| \ll 1$,
and we assume that there is CP conservation.
In that case,
there will be no phase differences in Eqs.~\ref{Eq do AziDun}
and Eq.~\ref{CPcons chi2} reads $\chi_2 = \eta_X \eta_{B_s} \eta_K = - 1$.
Under these assumptions,
we find
\begin{eqnarray}
& &
\frac{4 \Gamma_{B_s} |q_K|^2}{|A_{B^0_s \rightarrow \overline{K^0}}|^2}
\int_0^\infty \Gamma \left[ B^0_s \stackrel{t_B}{\rightarrow} K
\stackrel{t_K}{\rightarrow} f \right]\, dt_B =
\nonumber\\
& &
\hspace{25mm}
e^{- \gamma_S t_K} \left| A_{K_S \rightarrow f} \right|^2
(1 - y_s)
+ e^{- \gamma_L t_K} \left| A_{K_L \rightarrow f} \right|^2
(1 + y_s)
\nonumber\\
& &
\hspace{25mm}
- 2 e^{- \frac{\gamma_S + \gamma_L}{2} t_K}
\left| A_{K_S \rightarrow f} A_{K_L \rightarrow f} \right|
\sin{(\Delta m_K\, t_K - \phi)}
\frac{1}{x_s}\ .
\end{eqnarray}
Therefore,
as pointed out by Azimov and Dunietz,
the interference term is sensitive to $1/x_s$ \cite{AziDun}.

We now look at the cascade decays with
$B^0_s \rightarrow \overline{K^0} \pi^0$,
$B^0_s \rightarrow \overline{K^0} \rho^0$,
etc.
If these decays are dominated by the tree-level diagram,
then the phase of $\chi_2$ measures the CKM phase $\gamma$.
However,
the penguin diagram comes in with a different weak phase and,
moreover,
the tree level diagram is suppressed by color and by
$|V_{ub} V_{ud}/(V_{tb} V_{td})|$.
The tree level and the penguin diagrams could come into
$\chi_2$ in commensurate proportions,
with correspondingly large hadronic uncertainties.
As a result,
these decays may be of limited use.
Of course,
if we were to assume that there is CP conservation,
then the phase of $\chi_2$ would become well defined.
But,
in this case,
such an assumption is not warranted for it would just hide the large
theoretical errors that do exist in the determination of the phase
of $\chi_2$.

\subsection{Cascade decays $B \rightarrow D$}

Let us consider decays of the type
$B^0_d \rightarrow \{ D^0, \overline{D^0} \} \pi^0$,
$B_d \rightarrow \{ D^0, \overline{D^0} \} \rho^0$, etc.
Here,
the decay $B^0_d \rightarrow D^0 X$ is suppressed with respect to the
decay $B^0_d \rightarrow \overline{D^0} X$
by $\tilde{\epsilon}^\prime = |V_{ub} V_{cd}/(V_{cb} V_{ud})|$.
One might be tempted to conclude that everything follows as for the 
$B^+ \rightarrow \{ D^0, \overline{D^0} \} X^+$ decay chains,
and, in particular,
that one might also use these decays to get at $x_D$.
This is not the case,
because of the $B^0_d - \overline{B^0_d}$ mixing.
In order to understand this,
one should compare Fig.~2 and Fig.~4.
\begin{figure}
\centerline{\psfig{figure=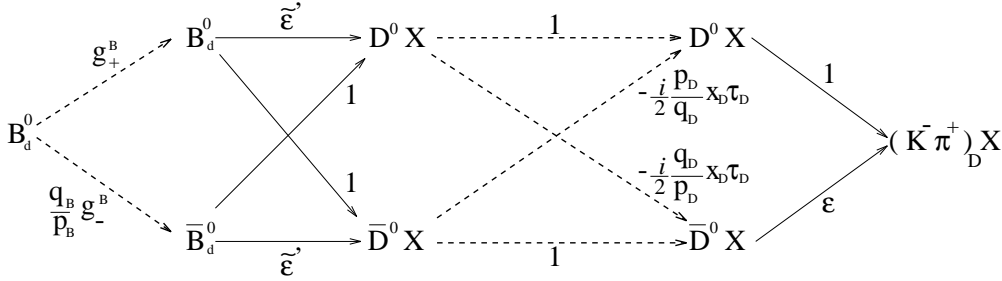,height=1.5in}}
\caption{The $B^0_d \rightarrow D X \rightarrow (K^- \pi^+)_D X$
cascade decay.
\label{fig:4}}
\end{figure}
The crucial point about Fig.~2 is that the two paths which exist
in the absence of $D^0 - \overline{D^0}$ mixing are both suppressed:
one by $\epsilon$, the other by $\tilde{\epsilon}$.
This is what makes the $x_D$ corrections important;
they are to be compared with $\epsilon$ and $\tilde{\epsilon}$,
and not with $1$ \cite{Silva}.
On the contrary,
in the $B_d \rightarrow D X$ decays of Fig.~4 there are four paths in the 
absence of $D^0 - \overline{D^0}$ mixing.
Two of the paths correspond to the unmixed $B^0_d \rightarrow B^0_d$
time evolution. One path is suppressed by $\epsilon$,
the other by $\tilde{\epsilon}^\prime$.
These paths are analogous to the ones in the $B^+ \rightarrow D X^+$
decay chain.
However,
in Fig.~4 there are also two paths corresponding to the mixed
evolution of $B^0_d$ into $\overline{B^0_d}$.
One of these is suppressed by $\epsilon \tilde{\epsilon}^\prime$,
but the other,
given schematically by
\begin{equation}
B^0_d \rightarrow \overline{B^0_d} \rightarrow D^0 X
\rightarrow D^0 X \rightarrow (K^- \pi^+)_D X,
\end{equation}
is not suppressed at all.
Therefore,
in $B_d \rightarrow D X$ decays,
the $x_D$ corrections are to be compared with $1$,
not with $\epsilon$ or $\tilde{\epsilon}^\prime$.
For $x_D \sim 10^{-2}$,
this correction amounts to about $1\%$,
and may be safely neglected.

There are, however,
two caveats to the analysis presented.
First,
if one is using the $B_d \rightarrow D X$ decays to look for precision
measurements,
such as effects proportional to $\epsilon$ or $\tilde{\epsilon}^\prime$,
then one must indeed take the $x_D$ corrections into account.
Second,
things would change if one could select events with
$t_B \ll 1/\Delta m_{B_d}$.
These times are too short for the $B^0_d$ to mix into
$\overline{B^0_d}$ and, for them,
the $B^0_d$ decay chain mimics that of a $B^+$.

We now turn to decays of the type
$B_s \rightarrow D X$,
where $X$ is a meson whose quark content contains a $s \bar{s}$ piece.
Examples include $\phi$, $\eta$, etc.
The analysis of these systems is exactly the same as for the previous
case,
except that the suppression of $\tilde{\epsilon}^\prime$ is
substituted here by a rather mild suppression given by 
$|V_{ub} V_{cs}/(V_{cb} V_{us})|= R_b$.
We have used $|V_{ub}| = R_b \lambda^3$,
with $R_b$ ranging from $1/3$ to $1/2$
to account for the fact that $|V_{ub}|$ is in fact closer to
$\lambda^3/2$ or even $\lambda^3/3$.
Decays of this type have been discussed by Gronau and London \cite{GL}.
As before,
the presence of mixing (here the $B^0_s - \overline{B^0_s}$ mixing)
means that the cascade decays based upon $B_s \rightarrow D \phi$
are much less affected by a large value of $x_D$ than the cascade
decays started by $B^+ \rightarrow D X^+$.

\subsubsection{The no-oscillation requirement}

While we may obtain values of $x_D \sim 10^{-2}$ in many models of new
physics,
the value of $x_D$ in the SM is much smaller than that.
One might wonder what happens to the $B \rightarrow D$ cascade decays if
we keep with the SM and take $x_D \sim y_D \sim 0$.
In this case there is no oscillation in the $D^0 - \overline{D^0}$ system:
$g_+^D = 1$ and $g_-^D = 0$.
The decay amplitude in Eq.~\ref{amplitude:P} becomes
\begin{equation}
\label{amplitude:P:B -> Dinto}
A \left[ B^0 \stackrel{t_B}{\rightarrow} D
\stackrel{t_D}{\rightarrow} [f]_D \right]
=
A_{B^0 \rightarrow D_{{\rm into\,  } f_D}}
\left[
g_+^B +
\lambda_{B \rightarrow D_{{\rm into\, } f_D}} g_-^B
\right]\ .
\end{equation}
In writing this expression we have used Eq.~\ref{chi3 e lambda}
and noticed that Eq.~\ref{state into f} implies
\begin{equation}
\label{decay into D into f}
A_{B^0 \rightarrow D_{{\rm into\, } f_D}}
\equiv \langle D_{{\rm into\, } f_D} | B^0 \rangle
=
A_{B^0 \rightarrow D^0} A_{D^0 \rightarrow f_D} +
A_{B^0 \rightarrow \overline{D^0}} A_{\overline{D^0} \rightarrow f_D}.
\end{equation}
Therefore,
in the absence of $D^0 - \overline{D^0}$ mixing,
the decay chain $B^0 \rightarrow D X \rightarrow [f]_D X$
reduces to the direct decay of an initial $B^0$ into the final state
$D_{{\rm into\, } f_D} X$.
This is as expected.
Indeed,
we know that $D_{{\rm into\, } f_D}$ is the combination of
$D^0$ and $\overline{D^0}$ that decays into the final state $[f]_D$
at time $t_D$.
But,
since there is no $D^0 - \overline{D^0}$ oscillation,
that is also the state that one must overlap with the state obtained
from the decay of the $B$ meson into the $D^0 - \overline{D^0}$ system
(at time $t_D=0$).

Typically,
$f_D$ is taken to be a flavour specific final state,
which picks up either $D^0$ or $\overline{D^0}$.
In fact,
we see from Eq.~\ref{state into f} that, in this case,
$|D_{{\rm into\, } f_D} \rangle \propto | D^0 \rangle$,
or $|D_{{\rm into\, } f_D} \rangle \propto | \overline{D^0} \rangle$.
Alternatively,
$f_D$ is taken to be a CP eigenstate, $f_{\rm cp}$, with eigenvalue
$e^{i \alpha_f} = \eta_f = \pm 1$.
In this context, it is usually assumed that there is no
CP violation in the decays of neutral $D$ mesons into the final
state $f_{\rm cp}$.
We can use Eqs.~\ref{need this} and \ref{state into f} to show that,
under these conditions,
${\cal CP} |D_{{\rm into\, } f_{\rm cp}} \rangle
= \eta_f |D_{{\rm into\, } f_{\rm cp}} \rangle$,
as expected.

This no-oscillation requirement is the one behind the
$B \rightarrow \{D^0, \overline{D^0}, D_{{\rm into\, } f_{\rm cp}}  \} X$
triangle construction in the Gronau-London-Wyler method to measure
the CKM phase $\gamma$ \cite{GL,GW}.

\subsection{Cascade decays
$B \rightarrow D_{{\rm into\, } f_D} K
\rightarrow D_{{\rm into\, } f_D} [f]_K$.}

Strictly speaking,
in the $B \rightarrow D K$ decay chains we must follow three time
dependences separately;
those of the evolution in the $B^0 -\overline{B^0}$,
$D^0 -\overline{D^0}$,
and $K^0 -\overline{K^0}$ systems.
However,
if we continue to assume that there is no evolution in the
$D^0 -\overline{D^0}$ system,
as we have done in Eq.~\ref{amplitude:P:B -> Dinto},
then these decay chains reduce to 
\begin{equation}
B \rightarrow D_{{\rm into\, } f_D} K \rightarrow 
D_{{\rm into\, } f_D} [f_K].
\end{equation}
This is effectively a $B \rightarrow K X \rightarrow [f]_K X$
decay chain of the type described by
Eqs.~\ref{amplitude: B -> K} and \ref{AziKay}.

There are two cases to be considered.
For $B_ d \rightarrow D K$ decay chains we need the amplitude
of $B^0_d \rightarrow D^0 K^0$,
which is proportional to $V_{ub}^\ast V_{cs}$,
and the amplitude of $B^0_d \rightarrow \overline{D^0} K^0$,
which is proportional to $V_{cb}^\ast V_{us}$.
These two amplitudes correspond to color-suppressed diagrams and differ
by about $R_b$.
They come into the amplitude of
$B^0_d \rightarrow D_{{\rm into\, } f_D} K$ in the proportions given by
Eq.~\ref{decay into D into f},
which depend on the final state $f_D$ into which
the $D^0 - \overline{D^0}$ combination decays.
We find,
\begin{eqnarray}
A_{B^0_d \rightarrow D_{{\rm into\, } f_D} K^0 }
& = &
A_{B^0_d \rightarrow \overline{D^0}\, K^0}
A_{\overline{D^0} \rightarrow f_D} +
A_{B^0_d \rightarrow D^0 K^0} A_{D^0 \rightarrow f_D}
\nonumber\\
& \sim &
\lambda^3 T^\prime A_{\overline{D^0} \rightarrow f_D}
+ R_b \lambda^3 e^{i \gamma} T A_{D^0 \rightarrow f_D},
\nonumber\\
A_{\overline{B^0_d} \rightarrow D_{{\rm into\, } f_D} \overline{K^0} }
& = &
A_{\overline{B^0_d} \rightarrow \overline{D^0}\, \overline{K^0}}
A_{\overline{D^0} \rightarrow f_D} +
A_{\overline{B^0_d} \rightarrow D^0 \overline{K^0}} A_{D^0 \rightarrow f_D}
\nonumber\\
& \sim &
R_b \lambda^3 e^{-i \gamma} T  A_{\overline{D^0} \rightarrow f_D}
+ \lambda^3 T^\prime A_{D^0 \rightarrow f_D},
\end{eqnarray}
where $T$ and $T^\prime$ contain the relevant form factors
and strong phases.
Gronau and London \cite{GL} have noticed that the two amplitudes involved
in these decays have roughly the same magnitude.
Using this fact they proposed a triangle construction based on the
comparison between
$|A_{B^0_d \rightarrow \overline{D^0} K^0}|$,
$|A_{B^0_d \rightarrow D^0 K^0}|$,
$|A_{B^0_d \rightarrow D_{{\rm into\, } f_{\rm cp}} K^0 }|$,
and the magnitudes of the CP conjugated decays.
This allows the extraction of the CKM phase $\gamma$,
up to discrete ambiguities.
Gronau and London \cite{GL} also noted that the interference CP-violating
terms in these decays allow the extraction of the unusual combination
of CKM phases $2 \beta + \gamma$.

The two amplitudes involved in $B_s \rightarrow D K$ decays
are those of $B^0_s \rightarrow D^0 \overline{K^0}$,
proportional to $V_{ub}^\ast V_{cd}$,
and $B^0_s \rightarrow \overline{D^0}\, \overline{K^0}$,
proportional to $V_{cb}^\ast V_{ud}$.
The former is CKM suppressed with respect to the latter by
roughly a factor of a few times $10^{-2}$.
This implies that triangles similar to those in the Gronau--London
construction \cite{GL} would be extremely narrow,
thus precluding a good measurement of $\gamma$ in this case.
In fact,
using Eq.~\ref{decay into D into f} we get
\begin{eqnarray}
\label{Eq do Mario}
A_{B^0_s \rightarrow D_{{\rm into\, } f_D} \overline{K^0} }
& = &
A_{B^0_s \rightarrow \overline{D^0}\, \overline{K^0}}
A_{\overline{D^0} \rightarrow f_D} +
A_{B^0_s \rightarrow D^0 \overline{K^0}} A_{D^0 \rightarrow f_D}
\nonumber\\
& \sim &
\lambda^2 T^\prime A_{\overline{D^0} \rightarrow f_D}
-
R_b \lambda^4 e^{i \gamma} T A_{D^0 \rightarrow f_D},
\nonumber\\
A_{\overline{B^0_s} \rightarrow D_{{\rm into\, } f_D} K^0 }
& = &
A_{\overline{B^0_s} \rightarrow \overline{D^0} K^0}
A_{\overline{D^0} \rightarrow f_D} +
A_{\overline{B^0_s} \rightarrow D^0 K^0} A_{D^0 \rightarrow f_D}
\nonumber\\
& \sim &
- R_b \lambda^4 e^{- i \gamma} T A_{\overline{D^0} \rightarrow f_D}
+ \lambda^2 T^\prime A_{D^0 \rightarrow f_D}.
\end{eqnarray}
However, this strong hierarchy has its advantages too.
Let us consider $f_D = f_{\rm cp}$ and assume that there is no
CP violation in the decays of neutral $D$ mesons into the final
state $f_{\rm cp}$.
In this case Eq.~\ref{CP inv with CP eigen} holds:
$| A_{M^0 \rightarrow f} | = 
| A_{\overline{M^0} \rightarrow f} |$.
By comparing  Eqs.~\ref{Eq do Mario} with Eqs.~\ref{Eq do AziDun}
we conclude that the proposal to measure $\Delta m_{B_s}$
put forth by Azimov and Dunietz \cite{AziDun},
can be replicated by using the cascade decay
$B_s \rightarrow D_{{\rm into\, } f_D} K \rightarrow
D_{{\rm into\, } f_D} [f_K]$ instead.
We recall that Eqs.~\ref{Eq do AziDun} imply that
$|\chi_2|$ may differ from $1$ by about $|P/T|$.
Here this problem is mitigated by the fact that
the second amplitude in the first Eq.~\ref{Eq do Mario} is suppressed by
$\sim 10^{-2}$.

\subsection{Cascade decays $D \rightarrow K$}

Let us now consider decays of the type
$D^0 \rightarrow \{ K^0, \overline{K^0} \}  X \rightarrow [f]_K X$.
These decays do not look promising.
We discuss them briefly, for the sake of completeness.
The decays involve the CKM matrix elements of the first two families.
In the standard phase convention for the CKM matrix,
these matrix elements are very approximately real.
Moreover, since $\delta_K \equiv |p_K|^2-|q_K|^2$ is known experimentally
to be very small,
the only rephasing-invariant quantities in
$D^0 \rightarrow \{ K^0, \overline{K^0} \}  X \rightarrow [f]_K X$
decays which have non-negligible weak phases are those involving
$q_D/p_D = e^{2 i \theta_D}$,
and this only if the new physics effects turn out to make
$\theta_D$ large.
Since the large weak phases ($\beta$ and $\gamma$) are not involved,
and since the kaon system is well tested experimentally,
these decays can only be used to obtain the mixing parameters in the
$D^0 - \overline{D^0}$ system.

\begin{figure}
\centerline{\psfig{figure=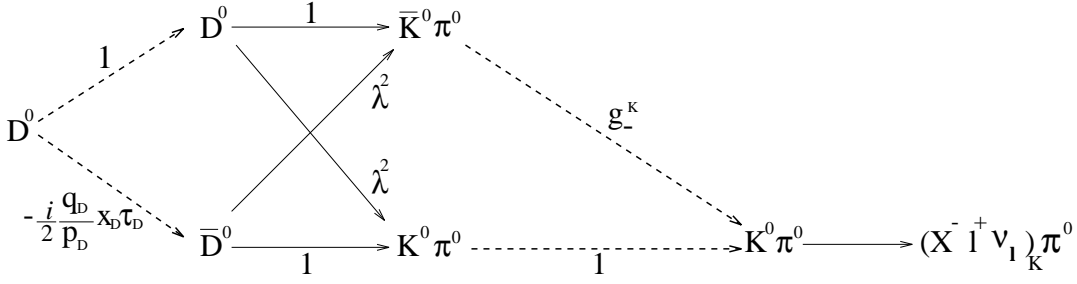,height=1.5in}}
\caption{The cascade decay
$D^0 \rightarrow K \pi^0 \rightarrow [X^- l^+ \nu_l]_K \pi^0$.
\label{fig:5}}
\end{figure}

However,
due to the fact that the mixing of the D meson is very small
(we continue to take $x_D \sim 10^{-2}$ as our reference value)
this mixing can only be relevant if the amplitudes
of the other paths available to get to the final state are suppressed in
some way
(as they are in the decay chain shown in Figs.~\ref{fig:2} and \ref{fig:3}).
This has two important consequences.
First,
one of the amplitudes,
$D^0 \rightarrow K^0 X$ or $D^0 \rightarrow \overline{K^0} X$,
must be suppressed relatively to the other.
The decays $D^0 \rightarrow K^0 \pi^0$ and
$D^0 \rightarrow \overline{K^0} \pi^0$ are of this kind,
since
\begin{equation}
|A_{D^0 \rightarrow K^0 \pi^0}| \sim \lambda^2
|A_{D^0 \rightarrow \overline{K^0} \pi^0}|.
\end{equation}
Second,
the final state $[f]_K$ cannot be a CP eigenstate,
since one of the $K \rightarrow [f]_K$
decay amplitudes must also be suppressed relatively to the other
(as in Fig.~\ref{fig:2}).
This second requirement is clearly satisfied by the semileptonic decays of
the neutral kaons, shown in Fig.~\ref{fig:5}.
In that case $A_{\overline{K^0} \rightarrow f} = 0$.

However,
the large mixing in the $K^0 - \overline{K^0}$ system will make it possible,
even then,
to get to the final state ($X^- l^+ \nu_l$) with no suppression at all.
This effect is similar to the one described in section 5.3 for the
mixing of the $B_d$.
So,
to measure the mixing parameters in the $D^0 - \overline{D^0}$ system
with this kind of decay,
we will also need to select events with very small $t_K$.
The idea is that,
for very short times $t_K$ the neutral kaons have not yet mixed appreciably,
and one might probe effects which are due mainly to the
mixing in the $D^0 - \overline{D^0}$ system.
But,
by looking at Fig.~\ref{fig:5} we recognize that this requires
a time cut of order $t_K \sim x_D/\Delta m_K$ or
$t_K \sim x_D/|\Delta \Gamma_K|$ for the $x_D$ effect to be dominant.
Experimentally,
$|\Delta \Gamma_K| \sim \gamma_S \sim 2 \Delta m_K$.
Therefore,
we would need to look at times of order $t_K \sim x_D \tau_S$,
where $\tau_S \sim 10^{-11} s$ is the lifetime of $K_S$!

This result can be checked directly in the decay rate.
Eq.~\ref{D-K} shows the expression for the decay rate,
integrated over $t_D$ and expanded only up to the terms linear in
$x_K \tau_K$ and $y_K \tau_K$.
Using Eq.~\ref{rate from P0}, we
find\footnote{It is
instructive to derive this formula in a different way.
Neglecting the weak phases of the first two families,
$\chi_3 = \chi_1 \chi_2$ and the 
$D^0 \rightarrow \{ K^0, \overline{K^0} \}  X \rightarrow [f]_K X$
decay rate may be written in a form similar to that of the
$B \rightarrow K$ decay rate shown in Eq.~\ref{AziKay}.
It is straightforward to use this to rederive Eq.~\ref{D-K}.}
%
\begin{eqnarray}
\label{D-K}
e^{\tau_K} \int_0^\infty \Gamma \left[ D^0 \stackrel{t_D}{\rightarrow} K
\stackrel{t_K}{\rightarrow} f \right]\, dt_D
&\propto&
1 + \mbox{Im} \chi_3 x_D
\nonumber\\
&+& y_K \tau_K \left[\mbox{Im}(\chi_1{\chi_3}^*) \frac{x_D}{2} -
  \mbox{Im}\chi_2 \frac{x_D}{2}-\mbox{Re}\chi_1\right]
\nonumber\\
&+& x_K \tau_K \left[\mbox{Im}\chi_1 +
  \mbox{Re}(\chi_1 {\chi_3}^*)\frac{x_D}{2} -
  \mbox{Re}\chi_2 \frac{x_D}{2} \right]
\end{eqnarray}
We have expanded in $x_D$ and taken $y_D=0$.
If we take the (mathematical) limit $\tau_K = 0$,
then we get a decay of the type already studied
by Wolfenstein \cite{Wol95},
with a term proportional to $\mbox{Im} \chi_3 x_D$.
However,
by developing Eq.~\ref{D-K} up to order $\tau_K^2$
and taking $\chi_3 = \chi_1 \chi_2$,
we find that the terms proportional to $x_D$ are only dominant
up to times of order $t_K \sim x_D\, \mbox{Re} \chi_2/\Delta m_K$,
or $t_K \sim x_D\, \mbox{Im} \chi_2/|\Delta \Gamma_K|$.
(Notice that $\mbox{Im} \chi_2$ can only be different from zero if the
$D^0 - \overline{D^0}$ mixing phase is also different from zero \cite{Wol95},
or if there is some strong phase involved \cite{Bla95,Bro95}.)
This confirms the result that we had guessed just by looking at
Fig.~\ref{fig:5}.
We conclude that, for times $t_K$ larger than about $x_D \tau_S$,
the effect of $x_D$ will be at most of order 1\%.

\section{Conclusions}

We have developed the formalism needed to study the most general cascade
decays of the type $P \rightarrow M + \cdots \rightarrow f + \cdots$,
which involve two neutral meson systems, $P$ and $M$, in succession.
The resulting decay rates exhibit the usual sources of CP violation:
CPV in the mixing of neutral meson systems,
probed by $|q/p| - 1$;
CPV present directly in the decays,
detected by
$|A_{i \rightarrow f}| - |A_{\bar i \rightarrow \bar f}|$;
and the CPV in the interference between the mixing in the initial neutral
meson system, $i$, and the decay {\em from that system}
into the final state $f$,
measured by
$\arg \lambda_{i \rightarrow f} + \arg \lambda_{i \rightarrow \bar f}$
(which is proportional to $\mbox{Im} \lambda_{i \rightarrow f}$ when $f$
is a CP eigenstate).

But,
when both the $P \rightarrow M + \cdots$ and
$P \rightarrow \overline{M} + \cdots$ decays are allowed,
we find a new class of rephasing-invariant parameters,
$\xi_i$,
that measure the interference between the mixing in the
$M^0 - \overline{M^0}$ system and the decay from the initial state
($P^0$ and $\overline{P^0}$) {\em into that system}.

We have applied this formalism to a variety of cases.
The main results are the following.
The proposal by Meca and Silva \cite{Silva}
to detect new physics in $x_D$ through the decays
$B^+ \rightarrow \{ D^0, \overline{D^0} \} K^+ \rightarrow [f]_D K^+$
may be extended to the
$B^+ \rightarrow \{ D^0, \overline{D^0} \} \pi^+$ and
$B^+ \rightarrow \{ D^0, \overline{D^0} \} \rho^+$ decays,
which have much larger branching ratios.
As we saw in Eqs.~\ref{not so obvious},
there are two interference effects that probe $x_D$.
If we neglect the strong phase differences,
one is proportional to a possible CP-violating phase in
the $D^0 - \overline{D^0}$ mixing due to new physics.
This effect might be easier to detect with the new decays proposed here
than it was in the original Meca and Silva proposal.
The other interference effect exists even in the absence of new
CP-violating phases in $D^0 - \overline{D^0}$ mixing,
and it involves the CKM phase $\gamma$.
When the branching ratios and suppression factors are taken into account,
this effect comes into the new decays at roughly the same
level as it does in $B^+ \rightarrow \{ D^0, \overline{D^0} \} K^+$
decays.

Next,
we showed that the mixing in the $B^0_d - \overline{B^0_d}$ system
implies that the same $x_D$ effects are much less important in the
$B_d \rightarrow \{ D^0, \overline{D^0} \} \pi^0$ and
$B_d \rightarrow \{ D^0, \overline{D^0} \} \rho^0$
(unless one could apply a very stringent time cut on the $B_d$ decays).
The same applies to $B_d \rightarrow \{ D^0, \overline{D^0} \} \phi$
decays.
Similarly,
$D \rightarrow \{ K^0, \overline{K^0} \} \pi^0$ decays cannot be used
to get a handle on $x_D$ unless one were able to perform a very
stringent time cut on $t_K$.

On the other hand,
if we neglect the $D^0 - \overline{D^0}$ mixing,
any $B \rightarrow D K \rightarrow [f]_D [f]_K$ decay chain 
(which would normally depend on three time variables, $t_B$, $t_D$, and $t_K$)
may be analyzed as a decay chain of the type 
$B \rightarrow D_{{\rm into}\ f_D} K \rightarrow D_{{\rm into}\ f_D} [f]_K$.
These chains can also be studied with the formalism developed in this article.
The most interesting case is $B_s \rightarrow  D_{{\rm into}\ f_{\rm cp}} K$.
In principle,
this could be used to get at $\Delta m_{B_s}$ in the
same way as proposed by Azimov and Dunietz with
the $B_s \rightarrow J/\psi K \rightarrow J/\psi [f]_K$ decay chains.

\vspace{5mm}

We thank C.\ Meca for discussions on related subjects,
and L.\ Lavoura for this and for reading and criticizing the manuscript.
We are indebted to B.\ Kayser for the seminars on cascade decays that
he gave in Lisbon, and which have inspired us to study this subject.
This work was partially supported by the Portuguese FCT under contract
Praxis/2/2.1/Fis/223/94 and by the grant CERN/P/FAE/1141/97.
The work of J.\ P.\ S.\ is partially supported by the grant
CERN/S/FAE/1164/97.
The work of M.\ G.\ Santos is partially supported by the
``Funda{\c c}{\~a}o da Universidade de Lisboa'', through CFNUL.

\vspace{5mm}

\appendix

\section{Time dependent functions involved in the decays of neutral meson
systems}
\label{appendix:evolution} 

Assuming CPT invariance,
the mass eigenstates of the $M^0 - \overline{M^0}$ system
are related to the flavour eigenstates by
\begin{eqnarray}
| M_H \rangle &=& p_M | M^0 \rangle + q_M | \overline{M^0} \rangle
\nonumber\\
| M_L \rangle &=& p_M | M^0 \rangle - q_M | \overline{M^0} \rangle
\label{eigenvectors}
\end{eqnarray}
with $|p_M|^2+|q_M|^2=1$ and
\begin{equation}
\frac{q_M}{p_M}
= \frac{\Delta m -\frac{i}{2} \Delta\Gamma}{2 R_{12}}=
\sqrt{\frac{R_{21}}{R_{12}}}
\end{equation}
where $\Delta m = m_H - m_L$ ($H$-heavy, $L$-light) is positive
by definition,
$\Delta\Gamma = \Gamma_H - \Gamma_L$,
and  $R_{12}$ is the off-diagonal matrix element in the effective
time evolution in the  $M^0 - \overline{M^0}$ space.

Consider a $M^0$ 
($\overline{M^0}$)
meson which is created at time $t_M=0$ and denote by $M^0(t_M)$
($\overline{M^0}(t_M)$)
the state that it evolves into after a time $t_M$,
measured in the rest frame of the meson M.
We find
\begin{eqnarray}
| M^0(t_M) \rangle &=& g_+(t_M) | M^0 \rangle +
\frac{q_M}{p_M} g_-(t_M) | \overline{M^0} \rangle,
\nonumber\\*[3mm]
| \overline{M^0}(t_M) \rangle &=& \frac{p_M}{q_M} g_-(t_M) | M^0 \rangle
+ g_+(t_M) | \overline{M^0} \rangle,
\label{timeevolution}
\end{eqnarray}
where
\begin{equation}
g_{\pm}(t) \equiv \frac{1}{2}
\left(
e^{-i \mu_H t} \pm e^{-i \mu_L t}
\right),
\end{equation}
$\mu_H \equiv m_H - i \Gamma_H/2$,
and $\mu_L \equiv m_L - i \Gamma_L/2$.
Similar expressions hold for the $P^0 - \overline{P^0}$ system.

The following formulas are useful:
\begin{eqnarray}
|g_\pm(t)|^2 & = &
\frac{1}{4} \left[
e^{- \Gamma_H t} + e^{- \Gamma_L t} \pm 2 e^{- \Gamma t} \cos(\Delta m\, t)
\right]
\label{gpmquad1}
\\*[3mm]
             & = &
\frac{e^{- \Gamma t}}{2}
\left[
\cosh \frac{\Delta \Gamma t}{2} \pm \cos \left( \Delta m t \right)
\right],
\label{gpmquad2}
\\*[3mm]
g_+^\ast(t) g_-(t) & = &
\frac{1}{4} \left[
e^{- \Gamma_H t} - e^{- \Gamma_L t} - 2 i e^{- \Gamma t} \sin(\Delta m\, t)
\right]
\label{g+*g-1}
\\*[3mm]
                   & = &
- \frac{e^{- \Gamma t}}{2} \left[
\sinh \frac{\Delta \Gamma t}{2} + i \sin \left( \Delta m t \right)
\right].
\label{g+*g-2}
\end{eqnarray}
Integrating over time,
we obtain
\begin{eqnarray}
G_\pm &\equiv& \int_0^{+ \infty} \!|g_\pm(t)|^2 dt =
\frac{1}{2 \Gamma} \left( \frac{1}{1 - y^2} \pm \frac{1}{1 + x^2} \right),
\label{Gpm}
\\*[3mm]
G_{+-} &\equiv& \int_0^{+ \infty} \!g_+^\ast(t) g_-(t) dt =
\frac{1}{2 \Gamma} \left( \frac{-y}{1 - y^2} + \frac{-ix}{1 + x^2} \right),
\label{G+-}
\end{eqnarray}
where $x \equiv \Delta m/\Gamma$ and $y = \Delta \Gamma/(2 \Gamma)$.


%

\begin{thebibliography}{99}
%
\bibitem{Chr64}
J.\ H.\ Christenson, J.\ W.\ Cronin, V.\ L.\ Fitch and R.\ Turlay,
Phys.\ Rev.\ Lett.\ {\bf 13}, 138 (1964).
%
\bibitem{CKM}
N.\ Cabibbo,
Phys.\ Rev.\ Lett.\ {\bf 10}, 531 (1963);
M.\ Kobayashi and T.\ Maskawa,
Prog.\ Theor.\ Phys.\ {\bf 49}, 652 (1973).
%
\bibitem{Sanda}
A.\ B.\ Carter and A.\ I.\ Sanda,
Phys.\ Rev.\ Lett.\ {\bf 45}, 952 (1980);
{\it ibid} Phys.\ Rev.\ D {\bf 23}, 1567 (1981);
I.\ I.\ Bigi and A.\ I.\ Sanda,
Nucl.\ Phys.\  {\bf B193}, 85 (1981);
{\it ibid} {\bf B281}, 41 (1987).
%
\bibitem{Azimov}
Ya.\ I.\ Azimov, JETP Letters {\bf 50}, 447 (1989);
Phys.\ Rev.\ D {\bf 42}, 3705 (1990).
%
\bibitem{Dass}
G.\ V.\ Dass and K.\ V.\ L.\ Sarma,
Int.\ J.\  Mod.\ Phys.\ {\bf A7}, 6081 (1992);
(E) {\it ibid} {\bf A8}, 1183 (1993).
%
\bibitem{Kay96}
B.\ Kayser and L.\ Stodolsky,
Max-Planck Institute preprint number MPI-PHT-96-112,
in hep-ph/9610522.
%
\bibitem{Kay97}
B.\ Kayser,
published in the proceedings {\it Les Arcs 1997, Electroweak
interactions and unified theories,} pages 389-398.
Also in hep-ph/9709382.
%
\bibitem{AziDun}
Ya.\ I.\ Azimov and I.\ Dunietz,
Phys.\ Lett.\ B {\bf 395}, 334 (1997).
%
\bibitem{Silva}
C.\ C.\ Meca and J.\ P.\ Silva,
CFNUL report number CFNUL/98-03,
in hep-ph/9807320.
To appear in Phys.\ Rev.\ Lett.\ (1998).
%
\bibitem{Luis}
G.\ C.\ Branco, W.\ Grimus and L.\ Lavoura,
Phys.\ Lett.\ B {\bf 372}, 311 (1996).
%
\bibitem{Dun95}
I.\ Dunietz,
Phys.\ Rev.\ D {\bf 52}, 3048 (1995).
%
\bibitem{Thank you}
We thank Ya.\ I.\ Azimov and I.\ Dunietz for
stressing the importance of this point,
in connection with their article in \cite{AziDun}.
%
\bibitem{Xing}
Z.-Z.\ Xing,
Phys.\ Lett.\ B {\bf 371}, 310 (1996).
%
\bibitem{ADS}
D.\ Atwood, I.\ Dunietz, and A.\ Soni,
Phys.\ Rev.\ Lett.\ {\bf 78}, 3257 (1997).
%
\bibitem{several}
For a review see, for example,
Y.\ Nir,
Nuovo Cim.\ {\bf 109A}, 991 (1996).
Generically,
these models keep $y_D \equiv \Delta \Gamma_D/(2 \Gamma_D)$ exceedingly
small,
and, thus, we will drop it from our analysis.
%
\bibitem{GL}
M.\ Gronau and D.\ London,
Phys.\ Lett.\ B {\bf 253}, 483 (1991).
%
\bibitem{GW}
M.\ Gronau and D.\ Wyler,
Phys.\ Lett.\ B {\bf 265}, 172 (1991).
%
\bibitem{Wol95}
L.\ Wolfenstein,
Phys.\ Rev.\ Lett.\ {\bf 75}, 2460 (1995).
See also
T.\ Liu,
Harvard University preprint number HUTP-94/E021,
in {\it Proceedings of the Charm 2000 Workshop},
FERMILAB-Conf-94/190, page 375, 1994.
%
\bibitem{Bla95}
G.\ Blaylock, A.\ Seiden, and Y.\ Nir,
Phys.\ Lett.\ B {\bf 355}, 555 (1995).
%
\bibitem{Bro95}
T.\ E.\ Browder and S.\ Pakvasa,
Phys.\ Lett.\ B {\bf 383}, 475 (1996).
%
\bibitem{PDG}
C.\ Caso {\it et al.},
European Physical Journal {\bf C3}, 1 (1998),
and also the URL:\ http://pdg.lbl.gov.
%
\bibitem{measure}
M.\ Athanas {\it et al.},
CLEO Collaboration,
CLEO preprint number CLEO 98-2.
%
\bibitem{Kayprivate}
B.\ Kayser,
private communication.
%
\end{thebibliography}
\end{document}